\documentclass[a4paper,11pt]{article}
\usepackage{jheppub}

\usepackage[utf8]{inputenc}
\usepackage{graphicx,cancel,float}
\usepackage{gensymb}
\usepackage{amssymb}
\usepackage{bbm}
\usepackage{textcomp}
\usepackage{amsmath}
\usepackage{tabularx}
\usepackage{bm}
\usepackage{times}
\usepackage{color}
\usepackage{slashed}
\usepackage{multirow}
\usepackage{verbatim}
\usepackage{enumerate}
\usepackage{epsfig,epstopdf}

\usepackage{indentfirst}
\usepackage{rotating}
\usepackage{subcaption}
\usepackage[mathscr]{eucal}
\usepackage{soul}

\allowdisplaybreaks[4]

\def\lsim{\mathrel{\raise.3ex\hbox{$<$\kern-.75em\lower1ex\hbox{$\sim$}}}}
\def\gsim{\mathrel{\raise.3ex\hbox{$>$\kern-.75em\lower1ex\hbox{$\sim$}}}}

\newcommand{\calO}{{\mathcal{O}}}

\definecolor{orange}{rgb}{1,0.5,0}

\title{Constraining super-light sterile neutrinos at Borexino and KamLAND}
\author[a]{Zikang Chen,}
\author[a]{Jiajun Liao,}
\author[a]{Jiajie Ling,}
\author[a]{Baobiao Yue}
\affiliation[a]{School of Physics, Sun Yat-Sen University, Guangzhou 510275, China }
\emailAdd{chenzk7@mail2.sysu.edu.cn}
\emailAdd{liaojiajun@mail.sysu.edu.cn}
\emailAdd{lingjj5@mail.sysu.edu.cn}
\emailAdd{yuebb@mail2.sysu.edu.cn}

\abstract{ 
The presence of a super-light sterile neutrino can lead to a dip in the survival probability of solar neutrinos, and explain the suppression of the upturn in the low energy solar neutrino data. In this work, we systematically study the survival probabilities in the 3+1 framework by taking into account of the non-adiabatic transitions and the coherence effect. We obtain an analytic equation that can predict the position of the dip. We also place constraints on the parameter space of sterile neutrinos by using the latest Borexino and KamLAND data. We find that the low and high energy neutrino data at Borexino are sensitive to different regions in the sterile neutrino parameter space. In the case with only $\theta_{01}$ being nonzero, the $\rm{{}^{8}B}$ data sets the strongest bounds at $\Delta m_{01}^{2} \approx (1.1\sim2.2)\Delta m_{21}^{2}$, while the low energy neutrino data is more sensitive to other mass-squared regions. The lowest bounds on $\Delta m_{01}^{2}$ from the $\rm{pp}$ data can reach $10^{-12} \ \rm{eV^{2}}$ because of the coherence effect. Also, due to the presence of non-adiabatic transitions, the bounds in the range of $10^{-9} \ \textrm{eV}^{2} \lesssim \Delta m_{01}^{2} \lesssim 10^{-5} \ \textrm{eV}^{2}$ become weaker as $\Delta m_{01}^{2}$ or $\sin^{2}2\theta_{01}$ decreases.  We also find that in the case with only $\theta_{02}$ or $\theta_{03}$ being nonzero,  the low energy solar neutrino data set similar but weaker bounds as compared to the case with only $\theta_{01}$ being nonzero. However, the bounds from the high energy solar data and the KamLAND data are largely affected by the sterile mixing angles.
}

\begin{document}

\maketitle
\setcounter{page}{2}

\newpage

\section{Introduction}
\label{sec:intro}
    Neutrino oscillation experiments provide a clear evidence of new physics beyond the Standard Model (SM). At present, most data from various neutrino oscillation experiments can be explained in the three neutrino oscillation framework with two mass-squared differences, $\Delta m_{21}^{2}\approx7.4\times 10^{-5} \ \rm{eV^{2}}$ and $\Delta m_{31}^{2}\approx2.5\times 10^{-3} \ \rm{eV^{2}}$~\cite{ParticleDataGroup:2020ssz}. However, the existence of sterile neutrinos is still a major open question in neutrino physics, and extensive searches for sterile neutrinos are under way; for recent reviews see Refs.~\cite{Abazajian:2012ys, Gariazzo:2015rra, Giunti:2019aiy, Boser:2019rta, Diaz:2019fwt, Dasgupta:2021ies, Hagstotz:2020ukm}. Currently, most of these searches focus on the eV-mass light sterile neutrinos, and are mainly motivated by anomalies in short baseline neutrino experiments at LSND~\cite{LSND:2001aii} and MiniBooNE~\cite{MiniBooNE:2018esg}, reactor neutrino experiments~\cite{Mueller:2011nm, Huber:2011wv} and Gallium experiments~\cite{Abdurashitov:2005tb}. Recently, very light sterile neutrinos are also studied to reconcile the tension between the NO$\nu$A and T2K data~\cite{deGouvea:2022kma}.
    In addition, super-light sterile neutrinos with a tiny mass splitting ($\lesssim \calO(10^{-5}) \ \rm{eV^{2}}$) against the active neutrinos also attract some interests in the literature~\cite{deHolanda:2010am,deHolanda:2003tx}.
    Super-light neutrinos can arise from a toy model with appropriate sterile couplings~\cite{Gomez-Izquierdo:2006bwq},
    a minimal radiative inverse seesaw model~\cite{BhupalDev:2012jvh} or quasi-Dirac scenarios~\cite{Donini:2011jh,Anamiati:2017rxw,Rossi-Torres:2013dya}. 
    The potential sensitivity to constrain super-light sterile neutrinos at JUNO and RENO-50 has been studied in Ref.~\cite{Bakhti:2013ora}. Limits on the quasi-Dirac neutrino parameter space have been derived from the solar, atmospheric neutrino data and cosmology~\cite{Cirelli:2004cz, deGouvea:2009fp}. 
    
    Current measurements of the smallest mass splitting among active neutrinos mainly come from the solar neutrino experiments and the medium-baseline reactor experiment at KamLAND~\cite{KamLAND:2008dgz}. However, a combined fit of $\Delta m_{21}^2$ from Super-Kamiokande and SNO data yields a small tension against the KamLAND data under the assumption of the CPT conservation~\cite{Super-Kamiokande:2016yck}.~\footnote{Note that this tension is reduced from about $2\sigma$ to $1.4\sigma$ with more statistics on the solar neutrino data; see the results released in Ref.~\cite{nu2020}.} This tension can be alleviated in the presence of a super-light sterile neutrino~\cite{deHolanda:2003tx,deHolanda:2010am}. 
    Due to the tiny mass splitting between the active and sterile neutrinos, the level-crossing and non-adiabatic transitions have to be taken into account during the propagation of solar neutrinos, and thus lead to a large modification to the solar neutrino survival probability. 
    It is also pointed out in Refs.~\cite{deHolanda:2010am,Cirelli:2004cz} that if the mass splitting between the sterile and active neutrinos is sufficiently small, the coherence bewtween different mass eigenstates plays an important role as solar neutrinos travel from the Sun to the Earth.

    In this work, we perform a systematical study of the propagation of solar neutrinos in the presence of a super-light sterile neutrino by taking into account of the non-adiabatic transitions and the coherence effect. In particular, we present a detailed analysis of the dependence of the survival probability on the neutrino energy, sterile mass-squared difference,  and mixing angle. 
    The Borexino experiment is located at the Laboratori Nazionali del Gran Sasso and utilizes the cleanest existing liquid scintillator detector to detect various components of solar neutrinos. Due to the extremely low background and good energy resolution, Borexino has successfully measured the $\rm{{}^{8}B}$~\cite{Borexino:2017uhp}, $\mathrm{pp}$~\cite{BOREXINO:2014pcl}, ${}^{7}\mathrm{Be}$~\cite{Bellini:2011rx,Borexino:2013zhu}, $\mathrm{pep}$~\cite{Borexino:2011ufb,Borexino:2017rsf}, as well as $\mathrm{CNO}$~\cite{BOREXINO:2020aww} neutrinos. In addition, 
    the KamLAND experiment provides the most precise determination of $\Delta m^{2}_{21}$ by measuring reactor neutrinos at Japan~\cite{KamLAND:2010fvi}. Here we also place constraints on the parameter space in the 3+1 framework using the current Borexino and KamLAND data. 
    
    This paper is organized as follows. In section~\ref{sec:prob}, we study the solar neutrino propagation in the presence of a super-light sterile neutrino. In section~\ref{sec:constraints}, we use the latest Borexino and KamLAND data to set constraints on the parameter space in the 3+1 framework. We summarize our results in section~\ref{sec:Con}.

\section{The formalism of solar neutrino propagation in the 3+1 framework}
\label{sec:prob}
We first describe the survival probability of solar neutrinos in the presence of a super-light sterile neutrino, then we discuss the impact on the survival probability from various sterile oscillation parameters.
\subsection{Survival probabilities in the 3+1 framework}
\label{sec:numerical}
    We consider the presence of a super-light sterile neutrino $\nu_{s}$ in addition to the three active neutrinos. We
    follow the notations in Refs.~\cite{deHolanda:2003tx,deHolanda:2010am}, and  denote the four mass eigenstates as $\nu_{0}$, $\nu_{1}$, $\nu_{2}$ and $\nu_{3}$. In the 3+1 framework, the evolution equation of solar neutrinos becomes
        \begin{equation}
            \label{eq.schr_4v}
            i \frac{d}{dx} |\nu_{\alpha}\rangle = H_{f} |\nu_{\alpha}\rangle \quad , \quad \alpha = s, e, \mu, \tau\,,
        \end{equation}
    with the Hamiltonian
        \begin{align}
        \label{eq.Hf_4v}
            H_{f}   &= U \mathrm{diag}(\frac{\Delta m_{01}^{2}}{2E_{\nu}}, 0, \frac{\Delta m_{21}^{2}}{2E_{\nu}}, \frac{\Delta m_{31}^{2}}{2E_{\nu}}) U^{\dagger} + V\,, 
          \end{align}  
  where $E_\nu$ is the neutrino energy, $\Delta m_{ij}^2$ are the mass-squared differences, 
  and the potential
          \begin{align}
            V       &= \mathrm{diag}(0, V_{CC}+V_{NC}, V_{NC}, V_{NC}) \\
                    &= \sqrt{2}G_{F}\mathrm{diag}(0, N_{e}-N_{n}/2, -N_{n}/2, -N_{n}/2)\,.
        \end{align}
  Here, $V_{CC}$ ($V_{NC}$) is the charged-current (neutral-current) potential, $N_{e}$ ($N_{n}$) is the number density of electron (neutron), and $G_{F}$ is the Fermi constant. Following Refs.~\cite{deHolanda:2010am,Bakhti:2013ora}, we parameterize the mixing matrix $U$ as
    \begin{eqnarray}
        \label{eq.U}
        U\equiv\left( \begin{matrix}1 & 0\cr 0 & U^{3\nu}\end{matrix}
        \right)\cdot U_S\,,
    \end{eqnarray}
    where $U^{3\nu}$ is the standard three-neutrino mixing matrix, i.e.,  the Pontecorvo-Maki-Nakagawa-Sakata (PMNS) mixing matrix, 
    \begin{equation}
    	\label{eq.Uv}
    	\begin{aligned}
    		U^{3\nu}    & = R_{23}(\theta_{23}) \cdot R_{13}(\theta_{13},\delta_{cp}) \cdot R_{12}(\theta_{12})\,, \\
    		& = \left(\begin{array}{ccc}
    			c_{13} c_{12} & c_{13} s_{12} & s_{13}e^{-i\delta_{cp}} \\
    			-c_{23}s_{12}-c_{12} s_{23} s_{13}e^{i\delta_{cp}} & c_{23} c_{12}-s_{23} s_{12} s_{13}e^{i\delta_{cp}} & c_{13} s_{23} \\
    			s_{23} s_{12}-c_{23} c_{12} s_{13}e^{i\delta_{cp}} & -c_{12} s_{23}-c_{23} s_{12} s_{13}e^{i\delta_{cp}} & c_{23} c_{13}
    		\end{array}\right)\,.
    	\end{aligned}
    \end{equation}
    Here, $R_{ij}$ is the rotation matrix with rotation angle $\theta_{ij}$ in the $ij$ plane, $s_{ij}$ ($c_{ij}$) denotes $\sin{\theta_{ij}}$ ($\cos{\theta_{ij}}$), and $\delta_{cp}$ represents the Dirac CP phase. The sterile mixing matrix
    $U_{S}$ is written as 
    \begin{eqnarray}
        \scriptsize
        U_{S}=
        \left(\begin{matrix}
        c_{01} & s_{01} e^{-i \delta_{01}} & 0 & 0 \\
        -s_{01} e^{i \delta_{01}} & c_{01} & 0 & 0 \\
        0 & 0 & 1 & 0 \\
        0 & 0 & 0 & 1
        \end{matrix}\right)
        \left(\begin{matrix}
        c_{02} & 0 & s_{02}e^{-i \delta_{02}} & 0 \\
        0 & 1 & 0 & 0 \\
        -s_{02}e^{i \delta_{02}} & 0 & c_{02} & 0 \\
        0 & 0 & 0 & 1
        \end{matrix}\right)
        \left(\begin{array}{cccc}
        c_{03} & 0 & 0 & s_{03} e^{-i \delta_{03}} \\
        0 & 1 & 0 & 0 \\
        0 & 0 & 1 & 0 \\
        -s_{03} e^{i \delta_{03}} & 0 & 0 & c_{03}
        \end{array}\right)\,.
    \end{eqnarray}
    
    
    In general, the survival probability of solar neutrinos $\nu_{e} \rightarrow \nu_{e}$ observed on the Earth can be written as
    \begin{equation}
        \label{eq:pee_nonad}
            P_{ee} = \left|\sum_{i=0}^{3}U_{ei}e^{-i\frac{\Delta m_{i1}^{2}}{2E_{\nu}}L_{0}}A_{ei}\right|^{2}\,,
    \end{equation}
    where $A_{ei}$ denotes the amplitude of transition $\nu_{e} \rightarrow \nu_{i}$ inside the Sun, and $L_{0} \simeq 1.5 \times 10^{11} \ \mathrm{m}$ is the distance between the Earth and the Sun. Here we ignore the Earth matter effects since the effect of day-night asymmetry and the conversion probability between the super-light sterile neutrino and active ones are at the level of a few percent on Earth~\cite{Liao:2014ola}. Hereinafter, we use the superscript '$M$' to represent the effective parameters in matter, with '$M_{0}$' referring specifically to those at the center of the Sun. 

    To study the effect of sterile mass-squared differences, we follow Ref.~\cite{deHolanda:2003tx,deHolanda:2010am}, and define $R_{\Delta} \equiv \Delta m_{01}^{2}/\Delta m_{21}^{2}$ for convenience. 
    We first consider the case in which the coherence effect can be ignored, and solar neutrinos arrived on the Earth can be treated as an incoherent sum of the four mass eigenstates. In this case, the survival probability in Eq.~(\ref{eq:pee_nonad}) can further simplified as 
     \begin{equation}
     \label{eq.pee_gen}
     P_{ee} = \sum_{i=0}^{3} \left|U_{e i}\right|^{2}\left|A_{ei}\right|^{2}\,.
     \end{equation}
     If we also assume that solar neutrinos propagate adiabatically, $A_{ei}$ is determined by elements of the effective mixing matrix in the center of the Sun, i.e., $A_{e i}=U_{e i}^{M_{0}}$, and we have
    \begin{equation}
    \label{eq.pee_ad_4v_l}
    P_{ee} = \sum_{i=0}^{3} \left|U_{e i}\right|^{2}\left|U_{ei}^{M_{0}}\right|^{2}\,.
    \end{equation}
    
    However, the adiabaticity of solar neutrino evolution can be violated in the presence of a super-light sterile neutrino. As we will show later that the level-crossing between two mass eigenstates has a large impact on the survival probability, and it is not feasible to use the adiabatic approximation in some cases. Also,  it is quite complicated to obtain an analytical solution for the hopping probability in the 3+1 case, we calculate the survival probability by solving the evolution equation in Eq.~(\ref{eq.schr_4v}) numerically. In our numerical procedure,
    we take the density profile of electrons and neutrons in the Sun from Ref.~\cite{Bahcall:2000nu}.  
    To solve the complex evolution equation in Eq.~(\ref{eq.schr_4v}), we first separate the imaginary part from the real part, and turn the 4-dimensional complex ordinary differential equations (ODE) into the 8-dimensional ODE; see Appendix~\ref{sec:Schrodinger} for details. By solving the 8-dimensional ODE with a numerical library for C and C++ programmers, GSL, we can obtain the amplitude of flavor transition $\nu_{e} \rightarrow \nu_{\alpha}$ inside the Sun, $\psi_{e\alpha}^{\rm{SS}}$. Here, the superscript 'SS' denotes the position at the surface of the Sun. Then, from Eq.~(\ref{eq:pee_nonad}), we can obtain the survival probability via 
    \begin{equation}
    	\label{eq.A_ei}
    	A_{ei}=\sum_{\alpha=s,e,\mu,\tau}U_{i\alpha}^{\dagger}\psi_{e\alpha}^{\rm{SS}}\,.
    \end{equation}

\begin{figure}[tbp]
    	\centering
    	\includegraphics[width=0.99\textwidth]{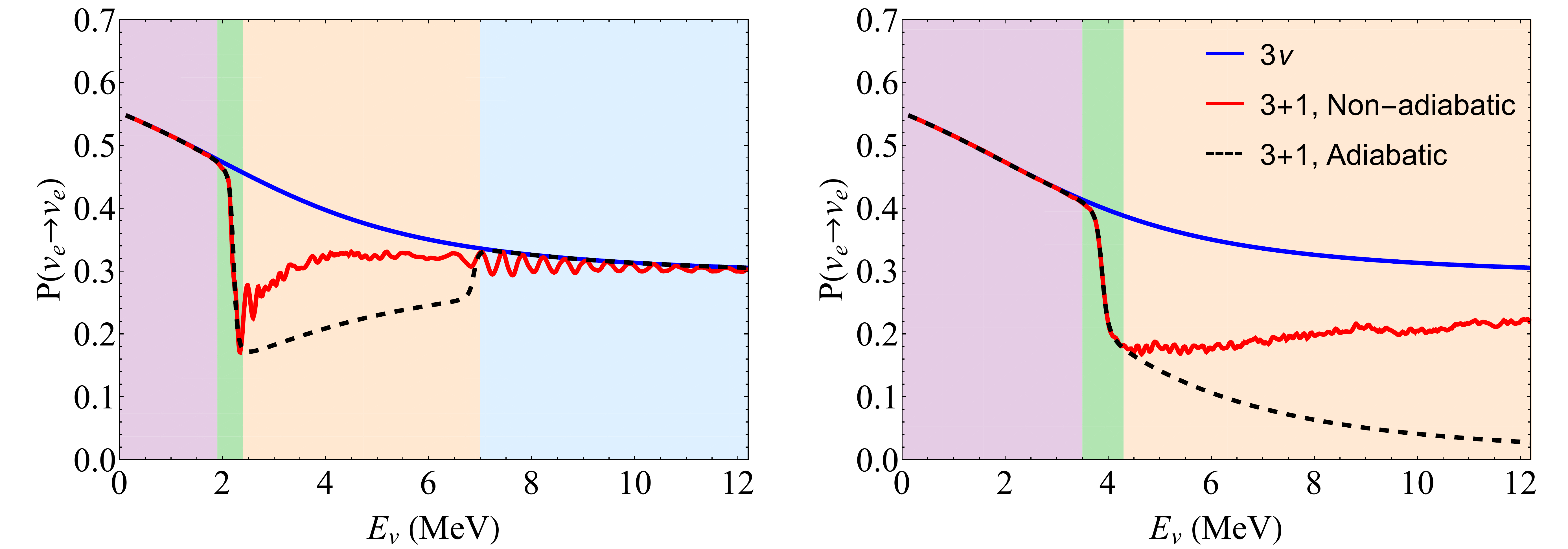}
    	\caption{Survival probabilities of solar neutrinos as a function of neutrino energy.  Here we take $\sin{^{2}2\theta_{01}} = 5\times10^{-4}$, and $R_{\Delta} = 0.15$ ($1.20$) in the left (right) panel. Other sterile neutrino parameters are set to be zero, and the oscillation parameters of active neutrinos are set at the best-fit values from the global fit in Ref.~\cite{2020}.
    	The red solid curves are obtained by using the numerical method in the 3+1 framework. The analytical solution in the 3+1 (3$\nu$) case under the assumption of adiabatic propagation is also shown as the black dashed (blue solid) curves for comparison. The colored regions are divided by different resonance energies; see the text for details.
    	}
    	\label{fig.Pee_AD_vs_NonAD}
    \end{figure}

    In the left and right panels of Fig.~\ref{fig.Pee_AD_vs_NonAD}, we show the survival probabilities in the 3+1 framework calculated with this numerical method for $R_\Delta=0.15$ and $R_\Delta=1.20$, respectively.
    For comparison, we also show the analytical solutions in the 3+1 and  3$\nu$ framework under the assumption of adiabatic propagation.
    From Fig.~\ref{fig.Pee_AD_vs_NonAD}, we see that for small sterile mixing angle, $P_{ee}$ in the 3+1 framework at low energies is consistent to $P_{ee}^{3\nu}$, and it can be described by the analytical solution in Eq.~(\ref{eq.pee_ad_4v_l}) under the assumption of adiabatic propagation. However, as $E_{\nu}$ increases, there is a large difference between the analytical solutions in Eq.~(\ref{eq.pee_ad_4v_l}) and the numerical results for $R_\Delta=0.15$  in the intermediate energy region and for $R_\Delta=1.20$  in the both the intermediate and high energy regions .
    
    \subsection{Level-crossing  and non-adiabatic effect}
    \label{sec:level-crossing}
    To understand the impact on the survival probabilities by the presence of a super-light sterile neutrino, we consider the case with $U_{S}=R_{01}(\theta_{01})$, i.e., $\nu_{s}$ only mixing with the mass eigenstate $\nu_{0}$ and $\nu_{1}$ for instance. 
    Following Ref.~\cite{deHolanda:2010am}, we rotate the Hamiltonian in Eq.~(\ref{eq.Hf_4v}) into a new basis, i.e., $|\tilde{\nu}\rangle = \tilde{U}^{\dagger} |\nu\rangle$ with $\tilde{U} = R_{23}R_{13}$. 
    The Hamiltonian in the new basis is
    \begin{equation}
    \label{eq.new_hf_v}
    \tilde{H_{f}}  = R_{12} R_{01} \mathrm{diag}(\frac{\Delta m_{01}^{2}}{2E_{\nu}}, 0, \frac{\Delta m_{21}^{2}}{2E_{\nu}}, \frac{\Delta m_{31}^{2}}{2E_{\nu}}) R_{01}^{\dagger} R_{12}^{\dagger} + \tilde{U}^{\dagger} V \tilde{U}\,.
    \end{equation}    
   Since $\Delta m_{31}^{2} \gg 2E_{\nu} V_{CC}$ is satisfied for solar neutrinos inside the Sun, we have  $\theta_{13}^{M} \approx \theta_{13}$, and the third mass eigenstate $\nu_3$ is decoupled from the other three mass eigenstates $\nu_0$, $\nu_1$ and $\nu_2$. 
   The solar neutrino evolution is dominantly governed by the 3 $\times$ 3 sub-matrix of $\tilde{H}_{f}$, which can be written as
    \begin{equation}
    \label{eq.H_new}
    (\tilde{H}_{f})_{3\times3} =  R_{12} R_{01} \mathrm{diag}(\frac{\Delta m_{01}^{2}}{2E_{\nu}}, 0, \frac{\Delta m_{21}^{2}}{2E_{\nu}}) R_{01}^{\dagger} R_{12}^{\dagger} + \tilde{V}\,,
    \end{equation}
    where
    \begin{equation}
    \tilde{V}    = \mathrm{diag}(0, V_{CC} \cos^{2} \theta_{13} + V_{NC}, + V_{NC})\,.
    \end{equation}
    Then, we rotate $(\tilde{H}_{f})_{3\times3}$ on the 1-2 plane with $R_{12}^{M}(\theta_{12}^{M})$, i.e.,
    \begin{equation}
    \label{eq.H_rotated}
    \begin{small}
    \begin{aligned}
    H_{M}   & = {R_{12}^{M}}^{\dagger} (\tilde{H}_{f})_{3\times3} R_{12}^{M} \\
    & \approx \left(\begin{array}{ccc}
    H_{00} & \ -\Delta m_{01}^{2}\sin{2\theta_{01}}\cos{(\theta_{12}-\theta^{M}_{12})}/4E_{\nu} & \ \Delta m_{01}^{2}\sin{2\theta_{01}}\sin{(\theta_{12}-\theta^{M}_{12})}/4E_{\nu} \\
    \cdots & \lambda_{1}^{\rm LMA} + \calO(\sin^{2}{\theta_{01}}) & 0 \\
    \cdots & \cdots & \lambda_{2}^{\rm LMA} + \calO(\sin^{2}{\theta_{01}})
    \end{array}\right)\,,
    \end{aligned}
    \end{small}
    \end{equation}
    where
    \begin{equation}
    \begin{footnotesize}
    \label{eq.d_th12_dx}
    \begin{aligned}
    \tan{2\theta_{12}^{M}} =\frac{(\Delta m_{21}^{2}-\Delta m_{01}^{2}\sin{^{2}\theta_{01}})\sin{2\theta_{12}}}{(\Delta m_{21}^{2}-\Delta m_{01}^{2}\sin{^{2}\theta_{01}})\cos{2\theta_{12}}-2E_{\nu}V_{CC}\cos{^{2}\theta_{13}}}\,.
    \end{aligned}
    \end{footnotesize}
    \end{equation}
    In the case of small sterile mixing angle $\theta_{01}$, the three eigenvalues of the Hamiltonian in Eq.~(\ref{eq.H_new}) can be written as
    \begin{equation}
    \label{eq.lamda_0}
     H_{00} = \frac{\Delta m_{01}^{2}\cos^{2}{\theta_{01}}}{2E_{\nu}}\,,
    \end{equation}
    \begin{equation}
    \begin{footnotesize}
    \label{eq.lamda_1}
    \begin{aligned}
    &\lambda_{1}^{\textrm{LMA}} = \frac{\Delta m_{21}^{2}}{4E_{\nu}} + \frac{2V_{NC}+V_{CC}\cos{\theta_{13}^{2}}}{2} - \sqrt{(\frac{\Delta m_{21}^{2}}{4E_{\nu}}\cos{2\theta_{12}}-\frac{V_{CC}\cos{\theta_{13}^{2}}}{2})^{2}+(\frac{\Delta m_{21}^{2}}{4E_{\nu}}\sin{2\theta_{12}})^{2}}\,,
    \end{aligned}
    \end{footnotesize}
    \end{equation}
    \begin{equation}
    \begin{footnotesize}
    \label{eq.lamda_2}
    \begin{aligned}
    &\lambda_{2}^{\textrm{LMA}} = \frac{\Delta m_{21}^{2}}{4E_{\nu}} + \frac{2V_{NC}+V_{CC}\cos{\theta_{13}^{2}}}{2} + \sqrt{(\frac{\Delta m_{21}^{2}}{4E_{\nu}}\cos{2\theta_{12}}-\frac{V_{CC}\cos{\theta_{13}^{2}}}{2})^{2}+(\frac{\Delta m_{21}^{2}}{4E_{\nu}}\sin{2\theta_{12}})^{2}}\,.
    \end{aligned}
    \end{footnotesize}
    \end{equation}
    Here, the superscript '$\rm{LMA}$' represents the large mixing angle (LMA) solution in the standard 3$\nu$ framework. From Eq.~(\ref{eq.d_th12_dx}), we can also get $\theta_{12}^{M} \approx \theta_{12}^{M,\rm{ LMA}}$ for small $\theta_{01}$. 

    \begin{figure}[htbp]
	\centering
	\includegraphics[width=0.99\textwidth]{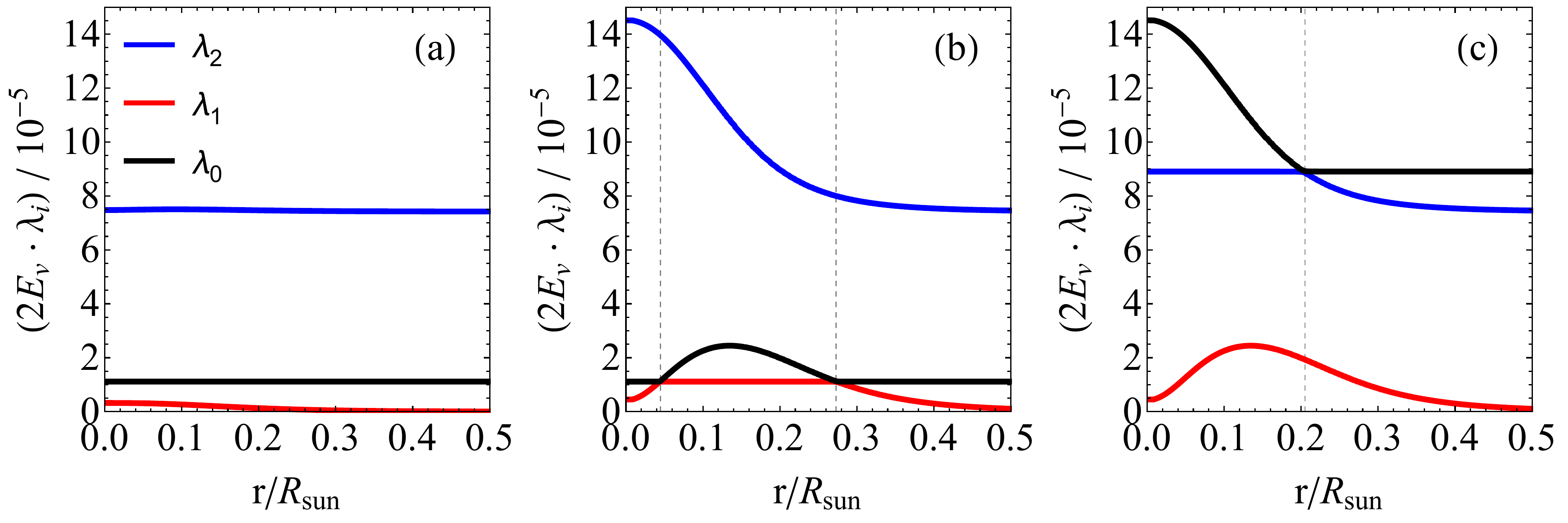}
	\caption{$2 E_\nu \lambda_{i}$ as a function of the propagation distance in the Sun. The left panel: (a) $R_{\Delta} = 0.15$, $E_{\nu}$ = 0.5 MeV; The middle panel: (b) $R_{\Delta} = 0.15$, $E_{\nu}$ = 10 MeV; The right panel: (c) $R_{\Delta} = 1.20$,  $E_{\nu}$ = 10 MeV. The mixing angle $\theta_{01}$ is fixed to be $\sin{^{2}2\theta_{01}} = 5\times10^{-4}$. Other oscillation parameters are the same as those in Fig.~\ref{fig.Pee_AD_vs_NonAD}.}
	\label{fig.cross}
\end{figure}

    We also introduce $\lambda_{i}$ to represent the sorted eigenvalues of $H_M$ for convenience, i.e., we take $\lambda_{2} > \lambda_{0} > \lambda_{1}$ ($\lambda_{0} > \lambda_{2} > \lambda_{1}$) for the case $R_{\Delta} < 1$ ($R_{\Delta} > 1$). The dependence of $\lambda_{i}$ on the propagation distance inside the Sun is shown in Fig.~\ref{fig.cross}. From Fig.~\ref{fig.cross}, we see that the level-crossing scheme inside the Sun is largely dependent on $E_{\nu}$ and $R_{\Delta}$. We can see from Fig.~\ref{fig.cross}(a) that for low energy neutrinos, since the matter effect can be ignored, the effective masses inside the Sun are close to the masses in vacuum and there is no level-crossing as neutrinos propagate through the Sun. For high energy neutrinos, if $R_{\Delta} = 0.15$ as shown in Fig.~\ref{fig.cross}(b), $\lambda_{0}$ can cross $\lambda_{1}$ twice during the propagation in the Sun. It is conceivable that, if $R_{\Delta}$ increases, the two resonance points in Fig.~\ref{fig.cross}(b) will get closer or even overlap. Inversely, if $R_{\Delta}$ decreases, the resonance point at low density will shift further away from the center of the Sun, while the one at high density becomes closer to the center of the Sun or even disappears, e.g., $\lambda_{0}$ cross $\lambda_{1}$ only once if the resonance point at high density disappears. From Fig.~\ref{fig.cross}(c), we can see that if $R_{\Delta} = 1.20$, $\lambda_{0}$ will always cross $\lambda_{2}$ once. Also, from Fig.~\ref{fig.cross}, we see that $\lambda_{1}$ and $\lambda_{2}$ never cross.
    
The active-sterile resonance energy can be determined by the resonance condition $\lambda_{0} = \lambda_{1}$ or $\lambda_{0} = \lambda_{2}$ . From Eqs.~(\ref{eq.lamda_0}), (\ref{eq.lamda_1}) and~(\ref{eq.lamda_2}), we can get
    \begin{equation}
    \label{eq.s_resonance}
    \begin{footnotesize}
    \begin{aligned}
    & E_{s} (x)= \frac{2\Delta m_{21}^{2}\cos{^{2}\theta_{01}}}{V_{CC}\cos^{2}\theta_{13}+2V_{NC}} \times \\
    & \frac{R_{\Delta}(1-R_{\Delta} \cos{^{2}\theta_{01}})}{(1-2R_{\Delta} \cos{^{2}\theta_{01}} + \xi \cos{2\theta_{12}})\pm\sqrt{(1-2R_{\Delta} \cos{^{2}\theta_{01}} + \xi \cos{2\theta_{12}})^{2}-4(\xi^{2}-1)R_{\Delta} \cos{^{2}\theta_{01}}(1-R_{\Delta} \cos{^{2}\theta_{01}})}}\,,
    \end{aligned}
    \end{footnotesize}
    \end{equation}
    where $\xi=V_{CC}\cos{^{2}\theta_{13}}/(V_{CC}\cos{^{2}\theta_{13}}-2V_{NC})$. Here the resonance energy $E_{s}$ is a function of the distance in the Sun. For details of the calculation of $E_{s}$; see Appendix.~\ref{sec:resonance_energy}. 
    As $E_{\nu}$ reaches $E_{s}$, the level-crossing between $\nu_{0}^{M}$ and $\nu_{1}^{M}$ or $\nu_{2}^{M}$ will occurs. From the discriminant in Eq.~(\ref{eq.delta}) of Appendix.~\ref{sec:resonance_energy}, we  know that if $R_{\Delta} \leq 0.2$ or $R_{\Delta} \geq 1$, the resonance energy at the center of the Sun $E_{s}(0)$ has two real roots that correspond to the solutions with '$+$' and '$-$' in Eq.~(\ref{eq.s_resonance}), and we denotes them as $E_{s}^{+}(0)$ and $E_{s}^{-}(0)$, respectively. For $0 < R_{\Delta} \leq 0.2$, $E_{s}^{+}(0)$ ($E_{s}^{-}(0)$) corresponds to the energy required for the first (second) level-crossing to occur at the high (low) density in the Sun, and we denote it as $E_{s1}$ ($E_{s2}$). In this case, if $E_{\nu}$ is large enough, the level-crossing between $\nu_{0}^{M}$ and $\nu_{1}^{M}$ in the Sun can be realized twice. For $R_{\Delta} > 1$, $E_{s}^{+}(0)$ is unphysical since it becomes negative, and $E_{s}^{-}(0)$ corresponds to $E_{s1}$. In this case, $\nu_{0}^{M}$ can cross $\nu_{2}^{M}$ only once in the Sun. For $R_{\Delta} < 0$, similar to the case $R_{\Delta} > 1$, $E_{s}^{+}(0)$ is also unphysical, and $E_{s}^{-}(0)$ corresponds to $E_{s1}$. However, $E_{s1}$ in this case is numerically larger than 20 MeV, which means that there is no level-crossing for solar neutrinos. If $0.2 < R_{\Delta} < 1$, $E_{s}(0)$ has no real root. For $E_\nu<20 \ \rm{MeV}$, we find that the $\lambda_{1}^{\rm{LMA}}$ in the Sun can reach $\sim0.4 \cdot \Delta m_{21}^{2}$ at most. Hence, for $0.2 < R_{\Delta} \lesssim 0.4$, $\nu_{0}^{M}$ can also cross $\nu_{1}^{M}$ twice in this case. Note that the level-crossing points do not occur at the center of the Sun in this case, as shown in Fig.~\ref{fig.cross}(b) for instance. For $0.4 \lesssim R_{\Delta} < 1$, $\nu_{0}^{M}$ can neither cross $\nu_{1}^{M}$ nor $\nu_{2}^{M}$ in the Sun, since the effective mass of $\nu_{0}^{M}$ in this case lies in the gap between the mass of $\nu_{1}^{M}$ and $\nu_{2}^{M}$.

Since the modifications to the survival probability are strongly dependent on the values of $R_{\Delta}$ and the neutrino energy, here we study the dependence of the survival probability on $E_{\nu}$ and $R_{\Delta}$ in more details. To illustrate the dependence on $E_{\nu}$, we divide the solar neutrino energies into different zones according to the resonance energies; see the colored regions in Fig.~\ref{fig.Pee_AD_vs_NonAD} for instance.
    
    \begin{enumerate}[(a)]
        \item $\mathbf{0 < R_{\Delta} \lesssim 0.4 }$
        
        In this case, solar neutrinos can pass through two resonance points at high energies as they propagate from the center to the surface of the Sun.

        \begin{enumerate}[i.]
            \item $\bf{E_{\nu} < E_{s1}:}$
                Since neutrino energy is small, the matter effect can be ignored in this zone, the adiabaticity of neutrino evolution remains in the Sun. Then we have $\left|A_{e i}\right|^{2}=\left|U_{e i}^{M_{0}}\right|^{2}$ and $P_{ee}$ can be obtained from Eq.~(\ref{eq.pee_ad_4v_l}).
                According to Eq.~(\ref{eq.U_01}) in Appendix.~\ref{sec:Um}, we get $\left|U_{e0}\right|^{2} \propto \sin^{2}{\theta_{01}} \sim \calO(10^{-3})$ is negligible, $\left|U_{e 2,3}\right|^{2}=\left|U_{e 2,3}^{3\nu}\right|^{2}$ and $\left|U_{e 1}\right|^{2}\approx\left|U_{e 1}^{3\nu}\right|^{2}$. We also find that $\left|U_{e 2,3}^{M_{0}}\right|^{2}=\left|U_{e 2,3}^{3\nu, M_{0}}\right|^{2}$ and $\left|U_{e 1}^{M_{0}}\right|^{2} = \left|U_{e1}^{3\nu, M_{0}}\right|^{2} - \left|U_{e 0}^{M_{0}}\right|^{2}$. Hence, the survival probability becomes
                \begin{equation}
                    \label{eq.pee_purple}
                    \begin{aligned}
                        P_{ee}  &\approx \left|U_{e 1}\right|^{2} \left(\left|U_{e 1}^{3\nu, M_{0}}\right|^{2}-\left|U_{e 0}^{M_{0}}\right|^{2}\right) + \left|U_{e 2}\right|^{2}\left|U_{e 2}^{M_{0}}\right|^{2} + \left|U_{e 3}\right|^{2} \left|U_{e 3}^{M_{0}}\right|^{2} \\
                                &\approx P_{ee}^{3\nu} - \left|U_{e 1}\right|^{2}\left|U_{e 0}^{M_{0}}\right|^{2}\,.
                    \end{aligned}
                \end{equation}
                Since $\theta_{01,12,13}^{M} \approx \theta_{01,12,13}$ in this case, $\left|U_{e0}^{M_{0}}\right|^{2} \approx \left|U_{e0}\right|^{2}$ can be also ignored. As a result, 
                \begin{equation}
                    \label{eq.pee_ad_3v}
                    P_{ee} \approx P_{ee}^{3\nu} \approx c^{4}_{13}(c^{4}_{12}+s^{4}_{12})+s^{4}_{13}\,,
                \end{equation}
                which can be seen from the purple zone in the left panel of Fig.~\ref{fig.Pee_AD_vs_NonAD}.
                
            \item $\bf{E_{\nu} \simeq E_{s1}} :$
                When $E_{\nu}$ is slightly larger than $E_{s1}$, $\theta_{01}^{M_{0}} \approx \frac{\pi}{2}$ in the center of the Sun. Hence, $\left|U_{e0}^{M_{0}}\right|^{2} \approx \left|U_{e1}^{3\nu,  M_{0}}\right|^{2}$. If the adiabaticity still maintains in this case,  Eq.~(\ref{eq.pee_purple}) becomes
                \begin{equation}
                    \label{eq.pee_green}
                        P_{e e} \approx P_{ee}^{3\nu} - \left|U_{e 1}\right|^{2}\left|U_{e 0}^{M_{0}}\right|^{2} \approx \left|U_{e 2}\right|^{2}\left|U_{e 2}^{M_{0}}\right|^{2} + \left|U_{e 3}\right|^{2} \left|U_{e 3}^{M_{0}}\right|^{2}\,.
                \end{equation}
                As we see form Eq.~(\ref{eq.pee_green}), the presence of $\left|U_{e 1}\right|^{2}\left|U_{e 0}^{M_{0}}\right|^{2}$ will trigger a dip in $P_{ee}$, as can be seen from the green zone in the left panel of Fig.~\ref{fig.Pee_AD_vs_NonAD}. In this case, the part of $\nu_{e}$ that would go to $\nu_{1}^{M}$ in the SM will be converted to $\nu_{0}^{M}$, and finally reaches the Earth in the form of $\nu_{0}$. For very small $\theta_{01}$, $\nu_{0}$ on the Earth is mainly converted to $\nu_{s}$ and can not be observed directly. Therefore, a significant dip in $P_{ee}$ occurs in the green energy region.
                
            \item $\bf{E_{s1} < E_{\nu} < E_{s2}}:$
                As $E_{\nu}$ increases, we take the non-adiabatic transition $\nu_{0}^{M} \leftrightarrow \nu_{1}^{M}$ into account, then 
                \begin{equation}
                    \label{eq.A}
                    A_{e0} = U_{e0}^{M_{0}} A_{00} + U_{e1}^{M_{0}} A_{10}, \quad A_{e1} = U_{e0}^{M_{0}} A_{01} + U_{e1}^{M_{0}} A_{11}\,.
                \end{equation}
                Inserting Eq.~(\ref{eq.A}) into Eq.~(\ref{eq.pee_gen}), we have
                \begin{equation}
                    \begin{aligned}
                    \label{eq.pee_resonance01_1}
                        P_{e e}
                        \approx& \left|U_{e 1}\right|^{2}\left|U_{e0}^{M_{0}} A_{01} + U_{e1}^{M_{0}} A_{11} \right|^{2} + \left|U_{e 2}\right|^{2}\left|U_{e 2}^{M_{0}}\right|^{2} + \left|U_{e 3}\right|^{2} \left|U_{e 3}^{M_{0}}\right|^{2} \\
                        \approx& P_{c}\left|U_{e 1}\right|^{2}\left|U_{e0}^{M_{0}}\right|^{2} + (1-P_{c})\left|U_{e 1}\right|^{2}\left|U_{e1}^{M_{0}}\right|^{2} + \left|U_{e 2}\right|^{2}\left|U_{e 2}^{M_{0}}\right|^{2} + \left|U_{e 3}\right|^{2} \left|U_{e 3}^{M_{0}}\right|^{2} \\
                        & + U_{e0}^{M_{0}}U_{e1}^{M_{0}}\left|U_{e 1}\right|^{2}\sqrt{P_{c}(1-P_{c})}\cos{\phi}\,,
                    \end{aligned}
                \end{equation}
                where $P_{c}\equiv\left|A_{ij}\right|^{2}$ is the hopping probability of the transition $\nu^{M}_i \rightarrow \nu^{M}_{j}$, and $\phi \equiv \rm{Arg}(A_{01}^{*}A_{11})$. As can be seen from the orange zone in Fig.~\ref{fig.Pee_AD_vs_NonAD}, due to the presence of $\phi$, there are some small wiggles appear in $P_{ee}$~\cite{deHolanda:2010am}. In addition, the dip within the orange zone is generally weakened due to the non-adiabatic transition $\nu_{0}^{M} \leftrightarrow \nu_{1}^{M}$. If the adiabaticity is strongly violated, i.e., $P_{c} \approx 1$, then we can get $P_{e e} \approx P_{ee}^{3\nu}$ from Eq.~(\ref{eq.pee_resonance01_1}).
                
            \item $\bf{E_{\nu} > E_{s2}}:$
                As the neutrino energy becomes larger than $E_{s2}$, neutrinos produced in the center of the Sun can pass through two non-adiabatic resonance points as they propagate through the Sun. In this case, $A_{01}$ and $A_{11}$ in Eq.~(\ref{eq.pee_resonance01_1}) need to be modified as
                \begin{equation}
                    \label{eq.Aij_1}
                        A_{01}  \rightarrow A_{00}^{h}A_{01}^{l} + A_{01}^{h}A_{11}^{l}\,, \\
                \end{equation}
                \begin{equation}
                    \label{eq.Aij_2}
                        A_{11}  \rightarrow A_{11}^{h}A_{11}^{l} + A_{10}^{h}A_{01}^{l}\,, \\
                \end{equation}
                where the superscripts '$h$' and '$l$' represent the resonance points at high and low density, respectively. It should be noted that $\theta_{01}^{M_{0}} \approx \theta_{01}$ in this case, so that $\left|U_{e 1}^{M_{0}}\right|^{2} \approx \left|U_{e 1}^{3\nu, M_{0}}\right|^{2}$ and $\left|U_{e 0}^{M_{0}}\right|^{2} \approx 0$. If the adiabaticity at both resonance points is strongly broken, namely $\left|A_{ij}^{h}\right|^{2} \approx \left|A_{ij}^{l}\right|^{2} \approx 1 (i \neq j)$ and $\left|A_{ii}^{h}\right|^{2} \approx \left|A_{ii}^{l}\right|^{2} \approx 0$, one can know from Eqs.~(\ref{eq.Aij_1}) and~(\ref{eq.Aij_2}) that $P_{c} = \left|A_{01}\right|^{2} \approx 0$. As a result, we can also get $P_{ee} \approx P_{ee}^{3\nu}$ from Eq.~(\ref{eq.pee_resonance01_1}).
                Note that $\theta_{12}^{M_{0}}$ approaches $\frac{\pi}{2}$ at high energies. In this case, $\nu_{e}$ produced in the center of the Sun basically consists of $\nu_{2}^{M}$ since $\left|U_{e2}^{M_{0}}\right|^{2} = \cos^{2}\theta_{13}^{M_{0}} \sin^{2}\theta_{12}^{M_{0}} \approx 1$, and $\nu_{2}^{M}$ adiabatically propagates to the surface of the Sun. As a result, the non-adiabatic transition $\nu_{0}^{M}\leftrightarrow \nu_{1}^{M}$ hardly has a impact on the survival probability, and $P_{ee} \approx \left|U_{e2}\right|^{2} = \cos^{2}\theta_{13}\sin^{2}\theta_{12} \approx 0.3$; see the blue zone in the left panel of Fig.~\ref{fig.Pee_AD_vs_NonAD}.
        \end{enumerate}
        
        \item $\mathbf{R_{\Delta} > 1}$
                
         In this case, there is at most one level-crossing point as solar neutrinos propagate from the center to the surface of the Sun.
        
        \begin{enumerate}[i.]
            \item $\bf{E_{\nu} < E_{s1}:}$ 
                At low energies, the matter effect is negligible, and the solar neutrino evolution is similar to the case with $0<R_{\Delta}\lesssim0.4$. 
                Since $\left|U_{e0}^{M}\right|^{2}$ and $\left|U_{e0}\right|^{2}$ are both small in this case, $P_{ee}$ described by Eq.~(\ref{eq.pee_ad_3v}) approaches the SM results; see the purple zone in the right panel of Fig.~\ref{fig.Pee_AD_vs_NonAD} for instance.
            
            \item $\bf{E_{\nu} \simeq E_{s1}:}$ 
                As we can see from the right panel of Fig.~\ref{fig.Pee_AD_vs_NonAD}, the probability in the green zone also has a big dip. Compare to the case with $0<R_{\Delta}\lesssim0.4$, it is that $\theta_{02}^{M_{0}} $ rather than $\theta_{01}^{M_{0}}$ approaches $\frac{\pi}{2}$ when $E_{\nu}$ is sightly higher than $E_{s1}$ in this case. According to Eq.~(\ref{eq.U_02}) in Appendix.~\ref{sec:Um}, we have $\left|U_{e1,3}\right|^{2}=\left|U_{e1,3}^{3\nu}\right|^{2}$, $\left|U_{e2}\right|^{2}\approx\left|U_{e2}^{3\nu}\right|^{2}$ and $\left|U_{e 0}^{M_{0}}\right|^{2} \approx \left|U_{e 2}^{3\nu, M_{0}}\right|^{2}$. If the adiabaticity of solar neutrino evolution still remains, the survival probability becomes
                \begin{equation}
                    \label{eq.pee_green_R1}
                    \begin{aligned}
                        P_{ee}  &\approx \left|U_{e 2}\right|^{2} \left(\left|U_{e 2}^{3\nu, M_{0}}\right|^{2}-\left|U_{e 0}^{M_{0}}\right|^{2}\right) + \left|U_{e 1}\right|^{2}\left|U_{e 1}^{M_{0}}\right|^{2} + \left|U_{e 3}\right|^{2} \left|U_{e 3}^{M_{0}}\right|^{2} \\
                                &\approx P_{ee}^{3\nu} - \left|U_{e 2}\right|^{2}\left|U_{e 0}^{M_{0}}\right|^{2}\,.
                    \end{aligned}
                \end{equation}
                 Similar to the case with $0<R_{\Delta}\lesssim0.4$, we can see that there is a dip caused by $\left|U_{e 2}\right|^{2}\left|U_{e 0}^{M_{0}}\right|^{2}$.

            \item $\bf{E_{\nu} > E_{s1}} :$
                In this case, $\nu_{0}^{M}$ has the level-crossing with $\nu_{2}^{M}$ rather than $\nu_{1}^{M}$, and we need to take the non-adiabatic transition $\nu_{0}^{M} \leftrightarrow \nu_{2}^{M}$ into account. With the exchange of indicators $1 \leftrightarrow 2$, $P_{ee}$ can be obtained from Eq.(\ref{eq.pee_resonance01_1}), i.e.,
                \begin{equation}
                    \begin{aligned}
                    \label{eq.pee_resonance02}
                        P_{e e}
                        \approx& \left|U_{e 2}\right|^{2}\left|U_{e0}^{M_{0}} A_{02} + U_{e2}^{M_{0}} A_{22} \right|^{2} + \left|U_{e 1}\right|^{2}\left|U_{e 1}^{M_{0}}\right|^{2} + \left|U_{e 3}\right|^{2} \left|U_{e 3}^{M_{0}}\right|^{2} \\
                        \approx& P_{c}\left|U_{e 2}\right|^{2}\left|U_{e0}^{M_{0}}\right|^{2} + (1-P_{c})\left|U_{e 2}\right|^{2}\left|U_{e2}^{M_{0}}\right|^{2} + \left|U_{e 1}\right|^{2}\left|U_{e 1}^{M_{0}}\right|^{2} + \left|U_{e 3}\right|^{2} \left|U_{e 3}^{M_{0}}\right|^{2} \\
                        & + U_{e0}^{M_{0}}U_{e1}^{M_{0}}\left|U_{e 2}\right|^{2}\cos{\phi}\sqrt{P_{c}(1-P_{c})}\,,
                    \end{aligned}
                \end{equation}
                where $\phi = \rm{Arg}(A_{02}^{*}A_{22})$. As we can see from the right panel of Fig.~\ref{fig.Pee_AD_vs_NonAD}, the dip in the orange zone still exists at high energies. The reason is that both $\theta_{01}^{M_{0}}$ and $\theta_{02}^{M_{0}}$ in this case are not equal to zero, and $U_{e2}^{M_{0}}$ is not simply proportional to $\sin\theta_{12}^{M_{0}}$; see Table.\ref{table.U} in Appendix.~\ref{sec:Um}. Unlike the case with $0<R_{\Delta}\lesssim0.4$, we find $\left|U_{e0}^{M_{0}}\right|^{2}\approx 1$ while $\left|U_{e2}^{M_{0}}\right|^{2}\approx 0$ at high energies in the case with $R_{\Delta}> 1$. It implies that $\nu_{e}$ produced in the center of the Sun basically consists of $\nu_{0}^{M_{0}}$, resulting in the dip in $P_{ee}$ at high energies.
        \end{enumerate}

      \end{enumerate} 
 
\begin{figure}[htbp]
	\centering
	\includegraphics[width=0.99\textwidth]{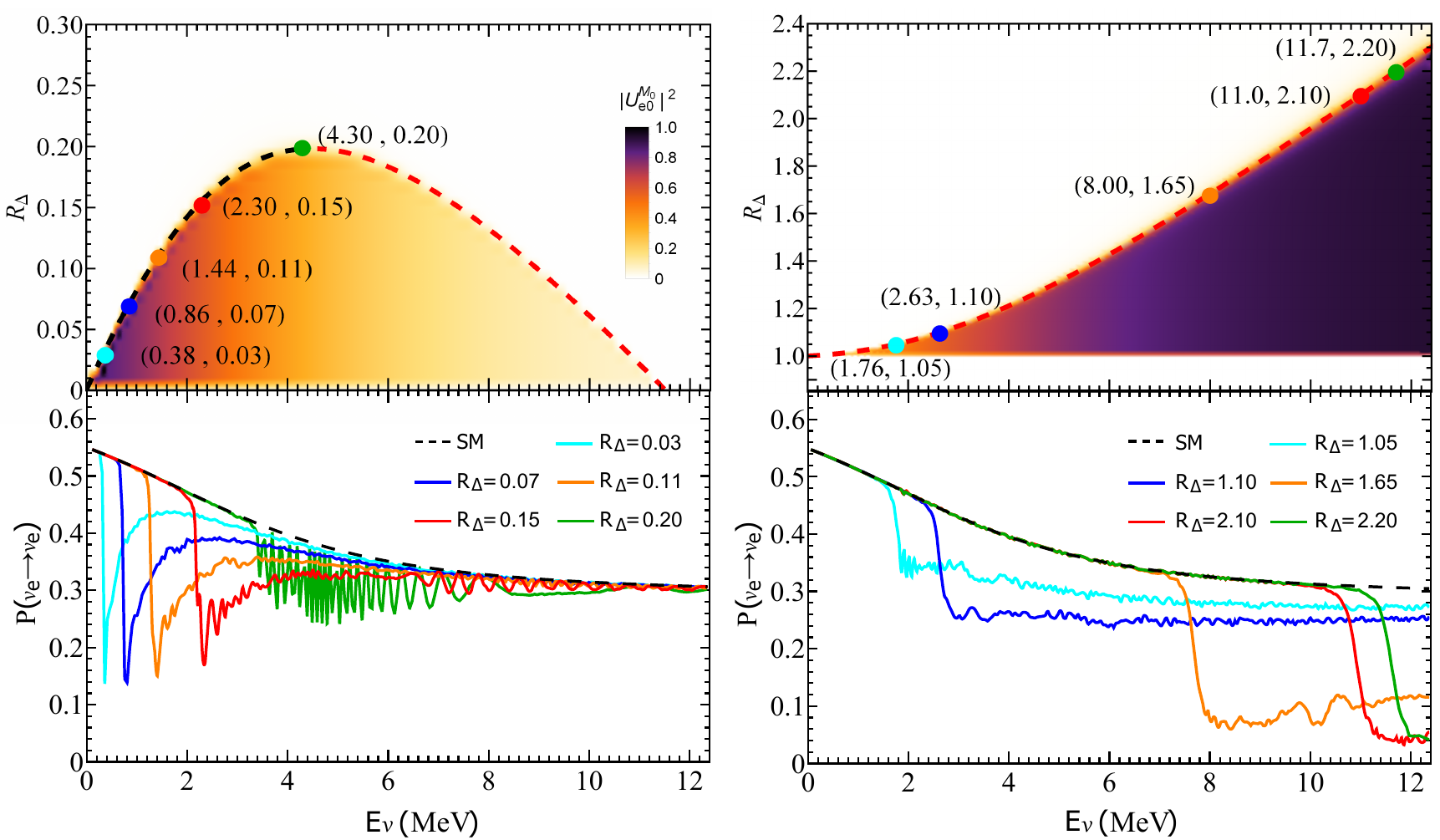}
	\caption{Upper panels: $\left|U_{e0}^{M}\right|^{2}$ as a function of neutrino energy and $R_{\Delta}$. The black (red) dashed curves correspond to $E_{s}^{+}(0)$ ($E_{s}^{-}(0)$) calculated by Eq.~(\ref{eq.s_resonance}). Lower panels: The survival probability as a function of neutrino energy for different $R_{\Delta}$. The left (right) panels correspond to $0<R_{\Delta}\lesssim0.4$ ($R_{\Delta}>1$). Here $\sin{^{2}2\theta_{01}}$ is set to be $5 \times 10^{-4}$, and other oscillation parameters not shown are the same as those in Fig.~\ref{fig.Pee_AD_vs_NonAD}. }
	\label{fig.Pee_dip}
\end{figure}
To further check the consistency of our analysis, we plot $\left|U_{e0}^{M_0}\right|^{2}$ and $P_{ee}$ for different values of $R_{\Delta}$ and $E_{\nu}$ with $\sin^{2}2\theta_{01} = 5 \times 10^{-4}$ using the numerical method.  
The results are shown in the left and right panels of Fig.~\ref{fig.Pee_dip} for $R_\Delta<1$ and $R_\Delta>1$, respectively. We also calculate $E_{s}^{+}(0)$ ($E_{s}^{-}(0)$) as a function of $R_\Delta$ by using the analytic solutions in Eq.~(\ref{eq.s_resonance}), which are shown as the black (red) dashed curves in the upper panels of Fig.~\ref{fig.Pee_dip}. We can see that the analytic results agree quite well with the boundaries of the color regions with $\left|U_{e0}^{M_0}\right|^{2}>0$. 
From the lower panels of Fig.~\ref{fig.Pee_dip}, we see that $P_{ee} \leq P_{ee}^{3\nu}$ for all neutrino energies in the presence of a super-light sterile neutrino. This can be understood as $|U_{e0}^{M_0}|$ can be large inside the Sun, it makes a part of $\nu_{e}$ leaving the Sun in the form of $\nu_{0}$, and finally results in the decrease of $P_{ee}$. 

From the lower panels of Fig.~\ref{fig.Pee_dip}, we see that there is a dip occurred at $E_\nu\simeq E_{s1}$ for each $P_{ee}$ curve, which can be explained by Eqs.~(\ref{eq.pee_green}) and~(\ref{eq.pee_green_R1}) for $R_\Delta<1$ and $R_\Delta>1$, respectively. 
In particular, we show the positions of the dip that occurred at the energies of $\rm{{}^{7}Be}$, $\rm{pep}$ and the peak energies of $\rm{pp}$ and $\rm{{}^{8}B}$. Compared to the upper panels of Fig.~\ref{fig.Pee_dip}, we find the positions of the dip in the left (right) panel are all located on the black (red) dash lines for a given $R_\Delta$. This demonstrates that the $E_{s}$ given by Eq.~(\ref{eq.s_resonance}) under different $R_{\Delta}$ is consistent with the resonance energy implied by $P_{ee}$ in Fig.~\ref{fig.Pee_dip}. Also, we find that the position of the dip in $P_{ee}$ shifts to a higher energy as $R_{\Delta}$ increases, which agrees with the trend of the dash lines. It is conceivable that for small mixing angle $\theta_{01}$, the position of the dip would be out of the energy range of solar neutrinos if $R_{\Delta}$ becomes too large. In this case, there is no effect on the survival probability from the conversion to sterile neutrinos.
 
\subsection{Dependence of the survival probability on the sterile mixing angles}
In this section, we study how the survival probability depends on the sterile mixing angles $\theta_{01}$, $\theta_{02}$ and $\theta_{03}$. In particular, we discuss the scenarios in which these sterile mixing angles are not small. Here for simplicity, we assume only one sterile mixing angle is nonzero at a time.

\subsubsection{Dependence on \texorpdfstring{$\theta_{01}$}{Lg}}
\label{subsubsec:theta_01}

\begin{figure}[tbp]
		\centering
		\setlength{\abovecaptionskip}{0.cm}
		\includegraphics[width=0.99\textwidth]{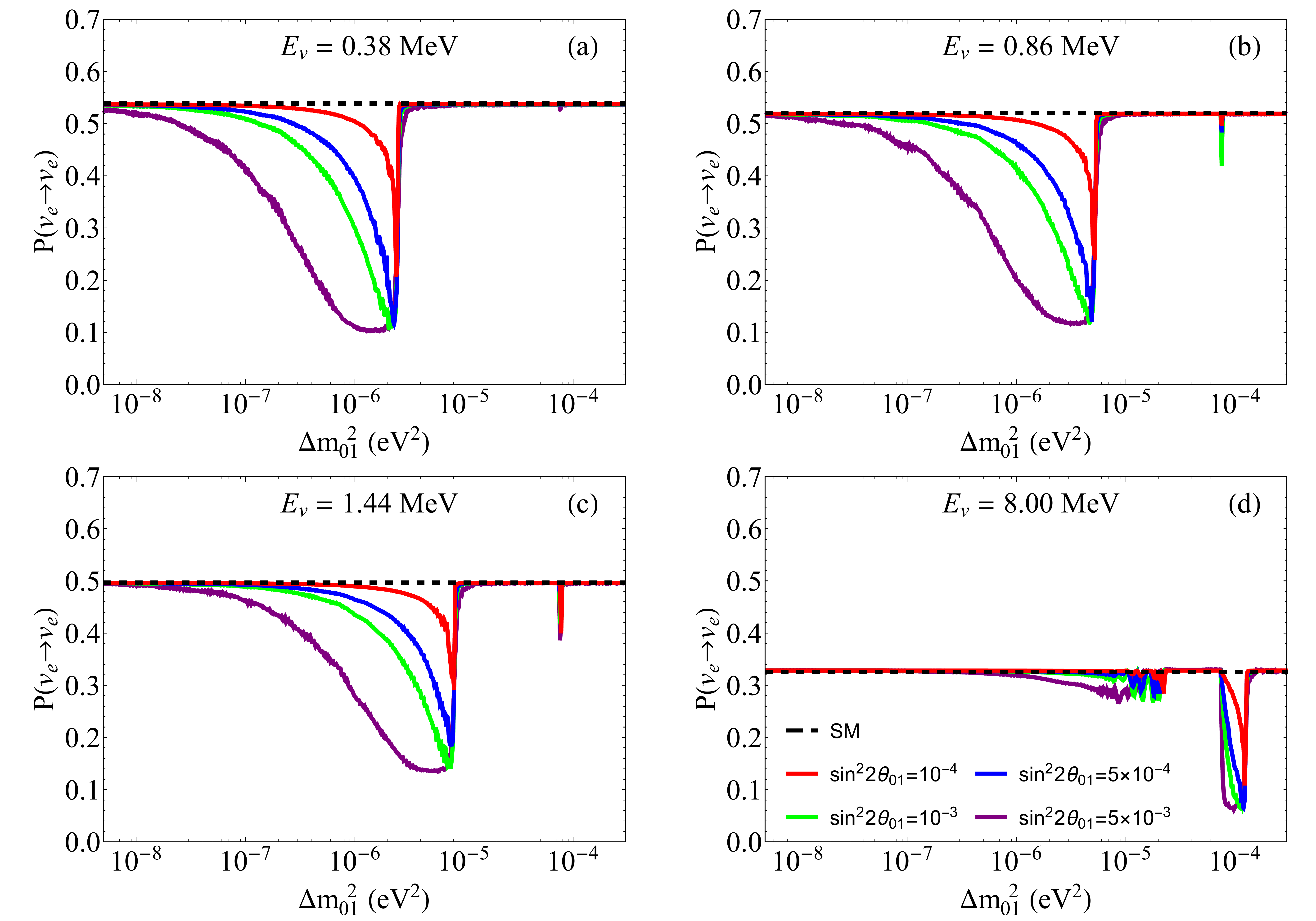}
		\caption{ Dependence of the survival probability on $\Delta m_{01}^{2}$ with different sterile mixing angle. The horizontal dashed line is the survival probability in the $3\nu$ case. The colored curves correspond to different value of $\sin{^{2}2\theta_{01}}$: $1\times10^{-4}$(Red), $5\times10^{-4}$(Blue), $1\times10^{-3}$(Green), and $5\times10^{-3}$(Purple), respectively. Other oscillation parameters not shown are the same as those in Fig.~\ref{fig.Pee_AD_vs_NonAD}.}
		\label{fig.Pee_vs_R}
	\end{figure}

Here we first focus on the case of $\nu_s$ only mixes in $\nu_{0}$ and $\nu_{1}$  in vacuum, i.e., only $\theta_{01}$ is nonzero for the sterile mixing angles.
%
	As shown in Fig.~\ref{fig.Pee_vs_R}, we obtain the survival probabilities as a function of $\Delta m_{01}^{2}$ for different values of the small mixing angle $\theta_{01}$. The neutrino energies are chosen to be 0.38 MeV, 0.86 MeV, 1.44 MeV, and 8.00 MeV, which correspond to the peak energies of $\rm{pp}$, $\rm{{}^{7}Be}$, $\rm{pep}$ and $\rm{{}^{8}B}$ neutrinos, respectively. From Fig.~\ref{fig.Pee_vs_R}, we can see that there are two very different dips. Among them, the first one is mainly located at $10^{-7} \ \rm{eV^{2}} \lesssim \Delta m_{01}^{2} \lesssim 10^{-5} \ \rm{eV^{2}}$, while the other one occurs when $\Delta m_{01}^{2} > \Delta m_{21}^{2}$. We can see that the first dips in Fig.~\ref{fig.Pee_vs_R}(a), (b) and (c) are much lower than the second dips. Since the low energy neutrinos predominantly propagate through the Sun in the $\nu_{1}^{M}$ state in the SM, the effect of the level crossing between $\nu_{0}^{M}-\nu_{1}^{M}$ on the survival probabilities is more significant than that for $\nu_{0}^{M}-\nu_{2}^{M}$. Moreover, the second dips in Fig.~\ref{fig.Pee_vs_R}(a), (b) and (c) are very narrow. That can be understood since the mass of $\nu_{2}^{M}$ hardly changes in the Sun at low energies; see Fig.~\ref{fig.cross}(a), so that $\nu_{0}^{M}$ can only cross $\nu_{2}^{M}$ within a very narrow range of $\Delta m_{01}^{2}$.
However, from Fig.~\ref{fig.Pee_vs_R}(d), we see the first dip is weakened while the second one becomes stronger at high energies. This is because most of solar neutrinos at high energies propagate through the Sun in the $\nu_2^{M}$ state, thus making the effect of the level crossing between $\nu_{0}^{M}-\nu_{2}^{M}$ more significant. Also, since the mass of $\nu_{2}^{M}$ varies in a large range at high energies; see Fig.~\ref{fig.cross}(c), the second dip in Fig.~\ref{fig.Pee_vs_R}(d) also becomes widened.
	Furthermore, from Fig.~\ref{fig.Pee_vs_R}, we find that the width of the dip is greatly influenced by $\sin^{2}2\theta_{01}$. The smaller the $\sin^{2}2\theta_{01}$ is, the narrower the width of the dip is. This is because the sterile mixing angle also plays an important role in non-adiabatic transitions between different mass eigenstates. By increasing $\sin^{2}2\theta_{01}$, one can effectively suppress the hopping probability of the non-adiabatic transitions, which leads to a wider dip in $P_{ee}$.
	
\begin{figure}[tbp]
	\centering
	\includegraphics[width=0.99\textwidth]{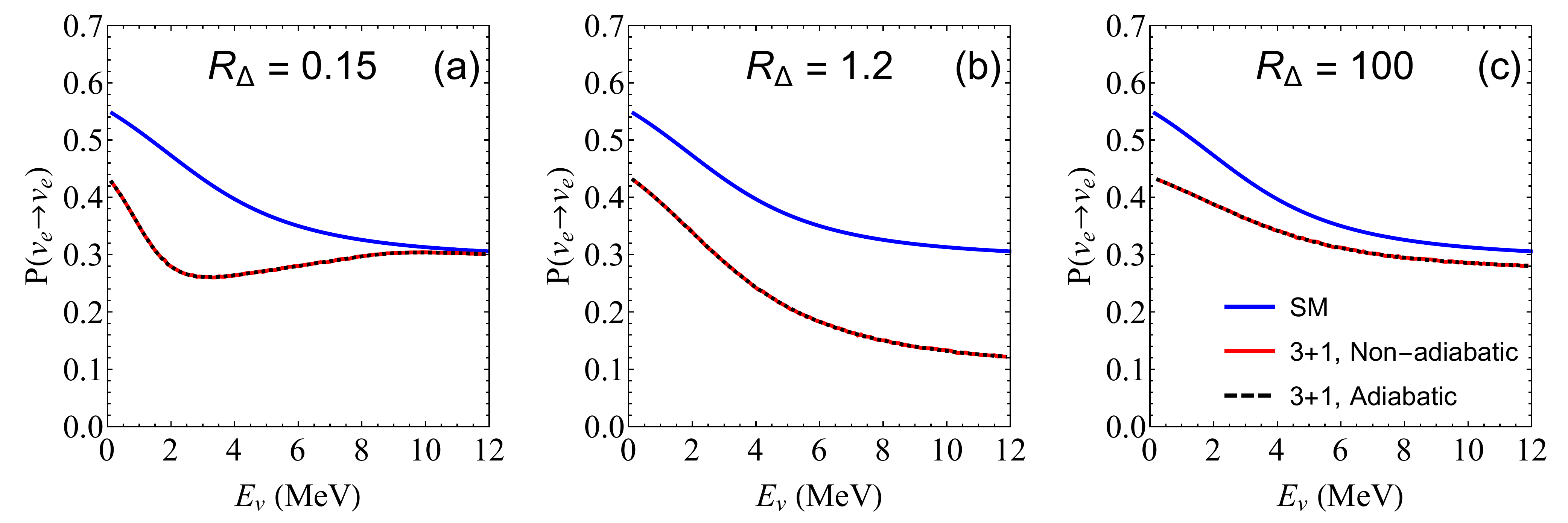}
	\caption{ Survival probabilities as a function of solar neutrino energy for $\sin{^{2}2\theta_{01}} = 0.5$. The solid blue and solid red lines correspond to numerical results in the $3\nu$ and 3+1 cases, respectively. The black dotted line is the analytical solution in the 3+1 case under the assumption of adiabatic propagation. Other oscillation parameters are the same as those in Fig.~\ref{fig.Pee_AD_vs_NonAD}.}
	\label{fig.posc_samples}
\end{figure}

    We also discuss the case in which the sterile mixing angle is not small, i.e., $\sin^{2}2\theta_{01}\sim \mathcal{O}(10^{-1})$.  In this case, $\left|U_{e2,3}\right|^{2} =\left|U_{e2,3}^{3\nu}\right|^{2}$, $\left|U_{e0}\right|^{2}=\sin^{2}\theta_{01}\left|U_{e1}^{3\nu}\right|^{2}$ and $\left|U_{e1}\right|^{2}=\cos^{2}\theta_{01}\left|U_{e1}^{3\nu}\right|^{2}$. For the neutrinos at low energies, we find that $\left|U_{ei}^{M_{0}}\right|^{2}$ has the similar relationship as $\left|U_{ei}\right|^{2}$. Hence, under the adiabatic approximation, $P_{ee}$ at low energies can be written as
	\begin{equation}
		\label{eq.big_m01_gg_m21}
		\begin{aligned}
		P_{e e}=& \sum_{i=0}^{3}\left|U_{e i}^{M_{0}}\right|^{2}\left|U_{e i}\right|^{2} \\
		=& P_{e e}^{3 \nu}-\frac{1}{2}\left(1-\cos 2 \theta_{01} \cos 2 \theta_{01}^{M_{0}}\right)\left|U_{e 1}^{3 \nu, M_{0}}\right|^{2}\left|U_{e 1}^{3 \nu}\right|^{2}\,.
		\end{aligned}
	\end{equation}
	For neutrinos at low energies, we have $\theta_{ij}^{M_{0}} \approx \theta_{ij}$($i\neq j$), then Eq.~(\ref{eq.big_m01_gg_m21}) can be simplified into 
	\begin{equation}
		\label{eq.big_th01_ad}
		P_{ee} \approx P_{e e}^{3 \nu}-\frac{1}{2}\sin^{2} 2 \theta_{01}\left|U_{e 1}^{3 \nu}\right|^{4}\,.
	\end{equation}
	From Eq.~(\ref{eq.big_th01_ad}), we know that there is a large difference between $P_{ee}$ and $P_{ee}^{3\nu}$ for large $\theta_{01}$. 
	We compare the above analytical and numerical results for a large sterile mixing angle and find that they are quite consistent with each other; see Fig.~\ref{fig.posc_samples} for instance. It demonstrates that the non-adiabatic transitions can be strongly suppressed and the adiabatic approximation still works for a large sterile mixing angle.

\begin{figure}[tbp]
	\centering
	\includegraphics[width=0.99\textwidth]{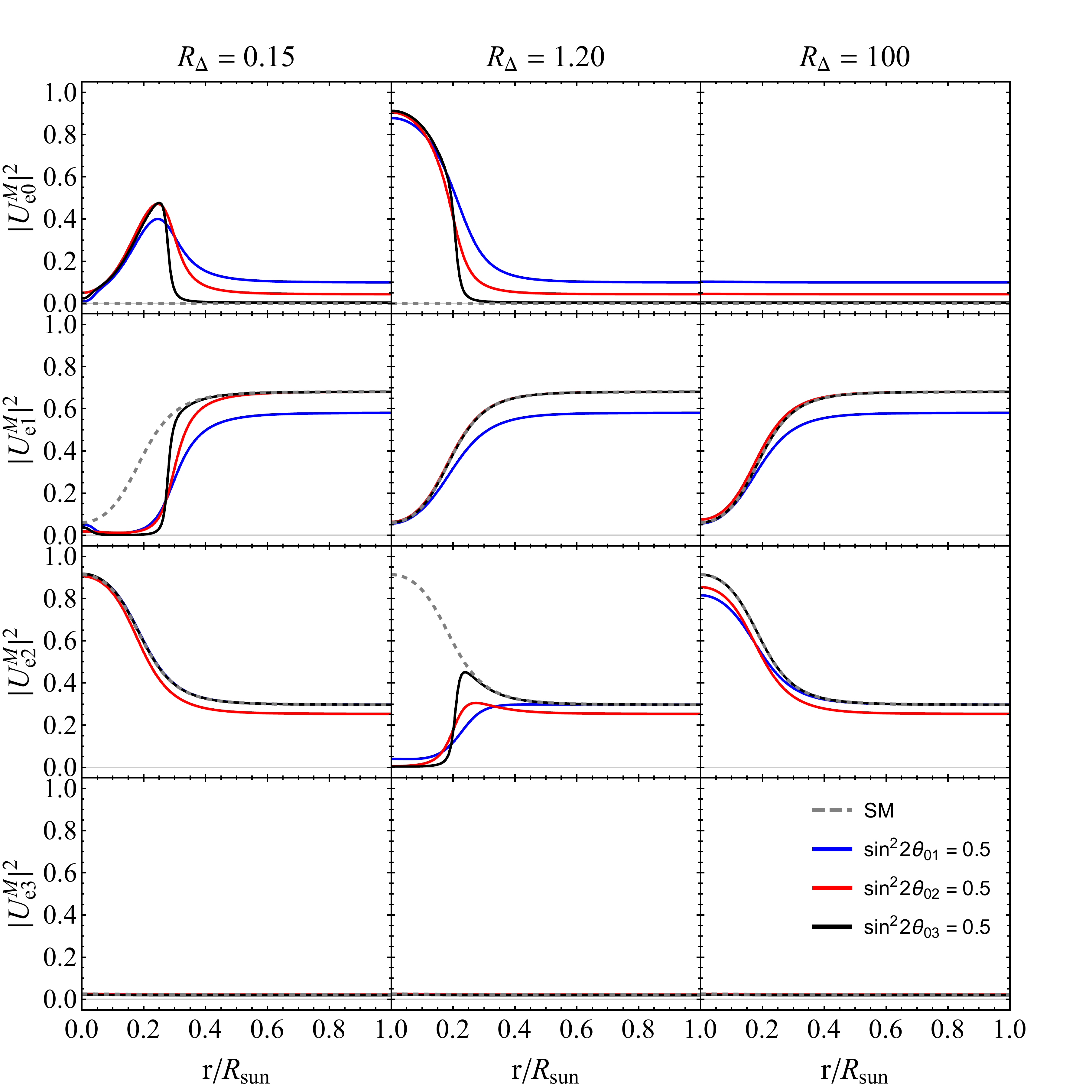}
	\caption{$\left|U_{ei}^{M}\right|^{2}$ as a function of the propagation distance in the Sun. The left (middle) [right] panels correspond to $R_{\Delta}=0.15$ (1.20) [100]. The neutrino energy is fixed to be $E_{\nu} = 10 \ \rm{MeV}$. Other oscillation parameters are the same as those in Fig.~\ref{fig.Pee_AD_vs_NonAD}.}
	\label{fig.U}
\end{figure} 
 	
	From Fig.~\ref{fig.posc_samples}, we see that unlike the case with a small sterile mixing angle, $P_{ee}$ at high energies is not the same for different values of $R_{\Delta}$. This can be understood by Fig.~\ref{fig.U}.
From Fig.~\ref{fig.U}, we see that $\left|U_{e2}^{M_{0}}\right|^{2}$ approaches 1 when $R_{\Delta} = 0.15$ or $R_{\Delta} = 100$, while $\left|U_{e0}^{M_{0}}\right|^{2}$ and $\left|U_{e1}^{M_{0}}\right|^{2}$ are very small. Hence, 
	\begin{align}
	    \label{eq.big_th01_R_l_1}
    	P_{ee} =\sum_{i=0}^{3}\left|U_{e i}^{M_{0}}\right|^{2}\left|U_{e i}\right|^{2}  
    	\approx \left|U_{e 2}\right|^{2} \approx \cos^{2}\theta_{13}\sin^{2}\theta_{12}\,.
	\end{align}
	From Eq.~(\ref{eq.big_th01_R_l_1}), we can see that $P_{ee}$ approaches $P_{ee}^{3\nu}$ at high energies; see Fig.~\ref{fig.posc_samples}(a). 
	However, for $R_{\Delta} = 1.20$, we find that it is $\left|U_{e 0}^{M_{0}}\right|^{2}$ that approaches 1 rather than $\left|U_{e2}^{M_{0}}\right|^{2}$. As a result,
    \begin{equation}
        \label{eq.big_th01_R_g_1}
        P_{ee} \approx \left|U_{e 0}\right|^{2} \approx \sin^{2}\theta_{01}\cos^{2}\theta_{13}\cos^{2}\theta_{12}\,.    
    \end{equation}
    As shown in Fig.~\ref{fig.posc_samples}(b), $P_{ee}$ can have a large deviation from $P_{ee}^{3\nu}$ at high energies. 

\subsubsection{Dependence on \texorpdfstring{$\theta_{02}$}{Lg} }
\label{sec.th02}
    Here we focus on the case $\nu_s$ only mixes in $\nu_{0}$ and $\nu_{2}$  in vacuum, i.e., only $\theta_{02}$ is nonzero in the sterile mixing angles.
        For the case of very small $\theta_{02}$, the effective Hamiltonian $H_{M}^{\prime}$ after rotation of $R_{12}^{M}$ is 
        \begin{equation}
            \label{eq.H_rotated_02}
            \begin{scriptsize}
                \begin{aligned}
                H_{M}^{\prime}  \approx \left(\begin{array}{ccc}
                                H_{00}+\calO(\sin^{2}\theta_{02}) & \ -\Delta m_{02}^{2}\sin{2\theta_{02}}\sin{(\theta_{12}-\theta^{M}_{12})}/4E_{\nu} & \ \Delta m_{02}^{2}\sin{2\theta_{02}}\cos{(\theta_{12}-\theta^{M}_{12})}/4E_{\nu} \\
                                \cdots & \lambda_{1}^{\rm {LMA}} + \calO(\sin^{2}{\theta_{02}}) & 0\\
                                \cdots & \cdots & \lambda_{2}^{\rm LMA} + \calO(\sin^{2}{\theta_{02}})
                                \end{array}\right) \,,
                \end{aligned}
            \end{scriptsize}
        \end{equation}
        which  can be obtained from Eq.~(\ref{eq.H_rotated}) by the following substitutions~\cite{deHolanda:2010am}:
        \begin{equation}
                \Delta m_{01}^{2} \rightarrow \Delta m_{02}^{2} \ , \ \theta_{01} \rightarrow \theta_{02}\,,
        \end{equation}
        \begin{equation}
                \sin{(\theta_{12}-\theta^{M}_{12})} \rightarrow -\cos{(\theta_{12}-\theta^{M}_{12})} \ , \ \cos{(\theta_{12}-\theta^{M}_{12})} \rightarrow \sin{(\theta_{12}-\theta^{M}_{12})}\,.
        \end{equation}
        If the sterile mixing angle $\theta_{02}$ is small, the diagonal terms in Eq.~(\ref{eq.H_rotated_02}) are approximately equal to those in Eq.~(\ref{eq.H_rotated}). Therefore, we can also get a similar expression for $E_{s}$ as in Eq.~(\ref{eq.s_resonance}), and the probabilities in this case are also similar to the case with only $\theta_{01}$ being nonzero.
        
        We also discuss the case with large $\theta_{02}$. In this case, $\left|U_{e1,3}\right|^{2} =\left|U_{e1,3}^{3\nu}\right|^{2}$, $\left|U_{e0}\right|^{2}=\sin^{2}\theta_{02}\left|U_{e2}^{3\nu}\right|^{2}$ and $\left|U_{e2}\right|^{2}=\cos^{2}\theta_{02}\left|U_{e2}^{3\nu}\right|^{2}$. For the neutrinos at low energies, $\left|U_{ei}^{M_{0}}\right|^{2}$ has the similar relationship as $\left|U_{ei}\right|^{2}$. Hence, under the adiabatic approximation, $P_{ee}$ at low energies is given by
        \begin{equation}
    		\label{eq.big_m01_gg_m21_02}
    		\begin{aligned}
    		P_{e e}= P_{e e}^{3 \nu}-\frac{1}{2}\left(1-\cos 2 \theta_{02} \cos 2 \theta_{02}^{M_{0}}\right)\left|U_{e 2}^{3 \nu, M_{0}}\right|^{2}\left|U_{e 2}^{3 \nu}\right|^{2}\,.
    		\end{aligned}
    	\end{equation}
        Since $\theta_{ij}^{M_{0}} \approx \theta_{ij}$($i\neq j$) is satisfied at low energies,  Eq.~(\ref{eq.big_m01_gg_m21_02}) becomes
        \begin{equation}
            \label{eq.big_th02_ad}
            P_{ee} \approx P_{e e}^{3 \nu}-\frac{1}{2}\sin^{2} 2 \theta_{02}\left|U_{e 2}^{3 \nu}\right|^{4}\,.
        \end{equation}
        Comparing with Eq.~(\ref{eq.big_th01_ad}), there is a less deviation from $P_{ee}^{3\nu}$ in Eq.~(\ref{eq.big_th02_ad}) due to $\left|U_{e 2}^{3 \nu}\right|^{4} < \left|U_{e 1}^{3 \nu}\right|^{4}$. For neutrinos at high energies, we find $\left|U_{e2}^{M_{0}}\right|^{2}$ approaches 1 if $R_{\Delta} = 0.15$ or $R_{\Delta} = 100$; see Fig.~\ref{fig.U}. Therefore, 
	\begin{align}
	    \label{eq.big_th02_R_l_1}
    	P_{ee} =\sum_{i=0}^{3}\left|U_{e i}^{M_{0}}\right|^{2}\left|U_{e i}\right|^{2}  
    	\approx \left|U_{e 2}\right|^{2} = \cos^{2}\theta_{02}\cos^{2}\theta_{13}\sin^{2}\theta_{12}\,.
	\end{align}
Similarly, if $R_{\Delta} =1.20$, we can see from Fig.~\ref{fig.U} that $\left|U_{e 0}^{M_{0}}\right|^{2}$ approaches 1 at high energies. Then, 
	\begin{align}
	    \label{eq.big_th02_R_g_1}
    	P_{ee} =\sum_{i=0}^{3}\left|U_{e i}^{M_{0}}\right|^{2}\left|U_{e i}\right|^{2}  
    	\approx \left|U_{e 0}\right|^{2} = \sin^{2}\theta_{02}\cos^{2}\theta_{13}\sin^{2}\theta_{12}\,.
	\end{align}
As we see from Eqs.~(\ref{eq.big_th02_R_l_1}) and~(\ref{eq.big_th02_R_g_1}), $P_{ee}$ becomes very different from $P_{ee}^{3\nu}$ for large $\theta_{02}$.

\subsubsection{Dependence on \texorpdfstring{$\theta_{03}$}{Lg}}
\label{sec:theta_03}
    Here we focus on the case $\nu_s$ only mixes in $\nu_{0}$ and $\nu_{3}$  in vacuum, i.e., only $\theta_{03}$ is nonzero in the sterile mixing angles. We firstly discuss the case of very small $\theta_{03}$. In this case, similar to Eq.~(\ref{eq.new_hf_v}), the solar neutrino evolution is dominantly governed by the 3 $\times$ 3 sub-matrix of $\tilde{H}_{f}^{\prime\prime}$, which can be written as
    \begin{equation}
    \label{eq.H_new_03}
    (\tilde{H}_{f}^{\prime\prime})_{3\times3} =  R_{12} R_{03} \mathrm{diag}(\frac{\Delta m_{01}^{2}}{2E_{\nu}}, 0, \frac{\Delta m_{21}^{2}}{2E_{\nu}}) R_{03}^{\dagger} R_{12}^{\dagger} + \tilde{V}\,.
    \end{equation}
After the rotation on the 1-2 plane with $R_{12}^{M}(\theta_{12}^{M})$, the Hamiltonian in Eq.~(\ref{eq.H_new_03}) becomes
\begin{equation}
    \label{eq.H_rotated_03}
    \begin{small}
    \begin{aligned}
    H_{M}^{\prime\prime}   & = {R_{12}^{M}}^{\dagger} (\tilde{H}_{f}^{\prime\prime})_{3\times3} R_{12}^{M} \\
    & \approx \left(\begin{array}{ccc}
    \frac{\Delta m_{01}^{2}}{2E_{\nu}}\cos^{2}\theta_{03}+\frac{\Delta m_{31}^{2}}{2E_{\nu}}\sin^{2}\theta_{03} & \ 0 & \ 0 \\
    \cdots & \lambda_{1}^{\rm{LMA}} & 0 \\
    \cdots & \cdots & \lambda_{2}^{\rm{LMA}}
    \end{array}\right)\,.
    \end{aligned}
    \end{small}
    \end{equation}

Since $\sin^{2}\theta_{03} \sim \calO(10^{-3})$ is negligible, $H_{M}^{\prime\prime}\approx \rm{diag(\frac{\Delta m_{01}^{2}}{2E_{\nu}}\cos^{2}\theta_{03}, \lambda_{1}^{\rm{LMA}}, \lambda_{2}^{\rm{LMA}})}$. From $\frac{\Delta m_{01}^{2}}{2E_{\nu}}\cos^{2}\theta_{03}=\lambda_{1}^{\rm{LMA}}$ or $\frac{\Delta m_{01}^{2}}{2E_{\nu}}\cos^{2}\theta_{03}=\lambda_{2}^{\rm{LMA}}$, we can get the same expressions for the resonance energy $E_{s}$ as Eq.~(\ref{eq.s_resonance}), except for replacing $\theta_{01}$ with $\theta_{03}$. Also, for the case of very small $\theta_{03}$, $\left|U_{ei}^{M}\right|^{2}$ are similar to those in the case of small $\theta_{01}$. Hence, the survival probabilities in this case have similar properties as those in the lower panels of Fig.~\ref{fig.Pee_dip}.

	In the case of large $\theta_{03}$, we have $\left|U_{e1,2}\right|^{2} =\left|U_{e1,2}^{3\nu}\right|^{2}$, $\left|U_{e0}\right|^{2}=\sin^{2}\theta_{03}\left|U_{e3}^{3\nu}\right|^{2}$ and $\left|U_{e3}\right|^{2}=\cos^{2}\theta_{03}\left|U_{e3}^{3\nu}\right|^{2}$. For the neutrinos at low energies, $\left|U_{ei}^{M_{0}}\right|^{2}$ has a similar relationship as $\left|U_{ei}\right|^{2}$. Hence, under the adiabatic approximation, we have
        \begin{equation}
    		\label{eq.big_m01_gg_m21_03}
    		\begin{aligned}
    		P_{e e}= P_{e e}^{3 \nu}-\frac{1}{2}\left(1-\cos 2 \theta_{03} \cos 2 \theta_{03}^{M_{0}}\right)\left|U_{e 3}^{3 \nu, M_{0}}\right|^{2}\left|U_{e 3}^{3 \nu}\right|^{2}\,.
    		\end{aligned}
    	\end{equation}
      If $\theta_{ij}^{M_{0}} \approx \theta_{ij}$($i\neq j$) is satisfied at low energies, then Eq.~(\ref{eq.big_m01_gg_m21_03}) becomes
        \begin{equation}
            \label{eq.big_th03_ad}
            P_{ee} \approx P_{e e}^{3 \nu}-\frac{1}{2}\sin^{2} 2 \theta_{03}\left|U_{e 3}^{3 \nu}\right|^{4}\,.
        \end{equation}
        Comparing with Eq.~(\ref{eq.big_th01_ad}), $P_{ee}$ in Eq.~(\ref{eq.big_th03_ad}) is approximately equal to $P_{ee}^{3\nu}$ since $\left|U_{e 3}^{3 \nu}\right|^{4}$ is negligible. For neutrinos at high energies, we find $\left|U_{e2}^{M_{0}}\right|^{2} \approx 1$ if $R_{\Delta} = 0.15$ or $R_{\Delta} = 100$; see Fig.~\ref{fig.U}. Therefore, $P_{ee}$ can be also described by Eq.~(\ref{eq.big_th01_R_l_1}) and approaches $P_{ee}^{3\nu}$.
However, if $R_{\Delta} =1.20$, we can see from Fig.~\ref{fig.U} that $\left|U_{e 0}^{M_{0}}\right|^{2}$ approaches 1 at high energies. Hence, 
	\begin{align}
	    \label{eq.big_th03_R_g_1}
    	P_{ee} =\sum_{i=0}^{3}\left|U_{e i}^{M_{0}}\right|^{2}\left|U_{e i}\right|^{2}  
    	\approx \left|U_{e 0}\right|^{2} = \sin^{2}\theta_{03}\sin^{2}\theta_{13}\,.
	\end{align}
Since $\sin^{2}\theta_{13}\approx 0.02$, one can get from Eq.~(\ref{eq.big_th03_R_g_1}) that $P_{ee} \approx 0.01$ for $\theta_{03}=\frac{\pi}{4}$.

\subsection{Coherence effect}
\label{sec:degenerate}
Due to the long distance between the Sun and the Earth, the coherence of the mass eigenstates is all averaged out for $\Delta m_{01}^{2} \gtrsim 10^{-9} \ \textrm{eV}^{2}$,  and neutrinos at the Earth can be treated as an incoherent sum of the mass eigenstates. However, if the mass-squared differences between $\nu_{0}$ and the other three mass eigenstates are small enough, the oscillation phase cannot be averaged out, and the coherence effect can have a large impact on the survival probabilities on Earth.
The oscillation phase of neutrino with a tiny mass split $\Delta m^{2}$ can be written as
\begin{equation}
\label{eq.phase}
\frac{\Delta m^{2} L_{0}}{4E_{\nu}} = 1.9\left(\frac{\Delta m^{2}}{10^{-11} \ \rm{eV^{2}}}\right)\left(\frac{1 \ \rm{MeV}}{E_{\nu}}\right)\,.
\end{equation}
Hence, if $\nu_{s}$ consists of a pair of quasi-degenerate states $\nu_{0}$ and $\nu_{1}$ with a tiny mass split $\Delta m^{2}_{01} \sim 10^{-11} \ \rm{  eV^{2}}$, the analytic formulas of oscillation probabilities for solar neutrinos
can be written as 
\begin{align}
P_{ee}      =& \left(1-\left|U_{e2}\right|^{2}-\left|U_{e3}\right|^{2}\right)\left(1-\left|U_{e2}^{M_{0}}\right|^{2}-\left|U_{e3}^{M_{0}}\right|^{2}\right) P_{ee}^{2f} \notag \\
& + \left|U_{e2}\right|^{2}\left|U_{e2}^{M_{0}}\right|^{2} + \left|U_{e3}\right|^{2}\left|U_{e3}^{M_{0}}\right|^{2}\,, \label{eq.pee_quasi_pee} \\
P_{es}      =& \left(1-\left|U_{e2}^{M_{0}}\right|^{2}-\left|U_{e3}^{M_{0}}\right|^{2}\right) \left(1-P_{ee}^{2f}\right)\,, \label{eq.pee_quasi_pes} 
\end{align}
where 
\begin{align}
P_{ee}^{2f} =& \sin^{2}\theta_{01} + \cos{2\theta_{01}}\left[P_{c}\sin^{2}\theta_{01}^{M_{0}}+\left(1-P_{c}\right)\cos^{2}\theta_{01}^{M_{0}}\right] \notag \\
&-\sqrt{P_{c}\left(1-P_{c}\right)}\cos{2\theta_{01}^{M_{0}}}\sin{2\theta_{01}}\cos{\left(2.54\frac{\Delta m_{01}^{2}}{E_{\nu}}L_{0}+\phi\right)}\,, \label{eq.pee_quasi_2f}
\end{align}
which is similar to the survival probability in the $2\nu$ case~\cite{deGouvea:2000pqg,deGouvea:1999wg}. Here $\phi$ is a phase caused by the matter effect and have a negligible contribution to the survival probabilities~\cite{deGouvea:2000pqg}. The hopping probability $P_{c}$ can be obtained by the expressions in the $2\nu$ case \cite{Pizzochero:1987fj,Kuo:1988pn}. At low energies, the matter effect can be ignored, i.e., $\left|U_{ei}^{M_{0}}\right|^{2}=\left|U_{ei}\right|^{2}$, then Eq.~(\ref{eq.pee_quasi_pee}) and Eq.~(\ref{eq.pee_quasi_pes}) become the same as Eq.~(IV.3) and Eq.(IV.4) in Ref.~\cite{deGouvea:2021ymm}, respectively.

\begin{figure}[tbp]
	\centering
	\setlength{\abovecaptionskip}{0.cm}
	\includegraphics[width=0.99\textwidth]{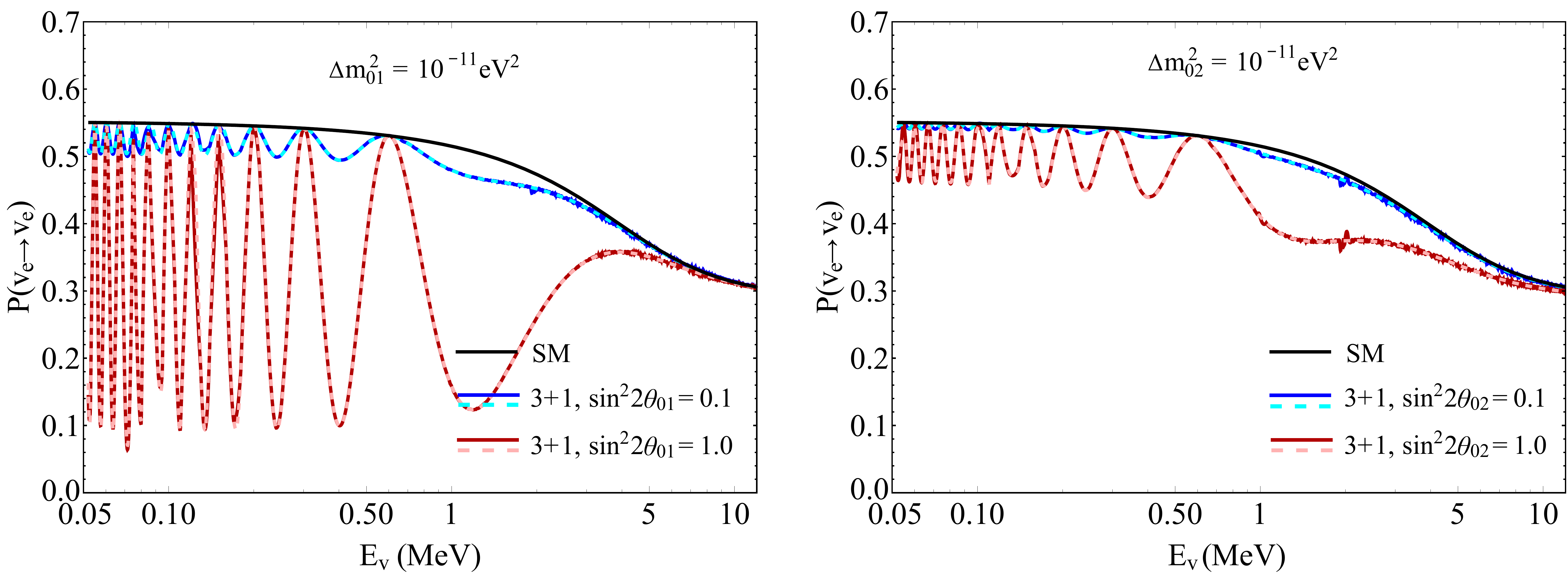}
	\caption{ Survival probability as a function of solar neutrino energy in the quasi-degenerate case. The dashed (solid) line represents the analytical (numerical) solution. $\Delta m_{01}^{2}(\Delta m_{02}^{2})$ is set to be $1 \times 10^{-11} \ \textrm{eV}^{2}$ in the left (right) panel. The blue and cyan (red and pink) curves correspond to $\sin^{2}2\theta_{01,02}= 0.1$ (1.0). Other oscillation parameters are the same as those in Fig.~\ref{fig.Pee_AD_vs_NonAD}.}
	\label{fig.Pee_coh_eng}
\end{figure}
In the left panel of Fig.~\ref{fig.Pee_coh_eng}, we show the survival probabilities that are calculated by Eq.~(\ref{eq.pee_quasi_pee}) and by the numerical method used in Sec.~\ref{sec:numerical}. We can see that the analytical solution agrees with the numerical results very well. In addition, it can be seen that the probabilities oscillate at low energies, which is caused by the cosine function in Eq.~(\ref{eq.pee_quasi_2f}). Since the amplitude of coherent oscillations is governed by $\sin2\theta_{01}$, $P_{ee}$ has a significant reduction in the low energy regions as $\sin2\theta_{01}$ increases. It implies that the low energy solar neutrinos will be sensitive to the quasi-degenerate states for a large sterile mixing angle. From Fig.~\ref{fig.Pee_coh_eng}, we can also see that as $E_{\nu}$ increases, $P_{ee}$ is consistent to $P_{ee}^{3\nu}$. The reason is that $\theta_{12}^{M_{0}} \approx \frac{\pi}{2}$ at high energies, then $P_{es} \approx 0$ and $P_{ee} \approx \left|U_{e2}\right|^{2} \approx 0.3$ due to $\left|U_{e2}^{M_{0}}\right|^{2} \approx 1$. It implies that the high energy solar neutrinos will not be sensitive to the quasi-degenerate states.

Also, if $\nu_{s}$ consists of a pair of quasi-degenerate states $\nu_{0}$ and $\nu_{2}$ with $\Delta m^{2}_{02} \sim 10^{-11} \ \rm{eV^{2}}$ and the sterile angle $\theta_{02}\sim\calO(10^{-1})$
one can obtain similar analytic formulas from Eqs.~(\ref{eq.pee_quasi_pee})~(\ref{eq.pee_quasi_pes}) and~(\ref{eq.pee_quasi_2f}) via the exchange of indicators $1\leftrightarrow2$. From the right panel of Fig.~\ref{fig.Pee_coh_eng}, we can see the coherence in this case has a weaker impact on the survival probabilities.

\section{Constraints from the experimental data}
\label{sec:constraints}
In this section, we present our results of constraints on the oscillation parameters of sterile neutrino using the experimental data from Borexino and KamLAND. 

\subsection{Experimental analysis}
    The Borexino experiment is located at the Laboratori Nazionali del Gran Sasso in Italy. The core of the detector is 278 ton of ultra-pure organic liquid scintillator, whose density of electrons is $N_{e} = (3.307 \pm 0.003)\times10^{31}/100$ ton~\cite{Borexino:2017rsf}. 
    Since the background can be greatly reduced by concentric layers of high purity materials, Borexino has the ability to measure the low-energy solar neutrinos $\rm{pp}$, $\rm{pep}$, $\rm{{}^{7}Be}$, and $\rm{CNO}$. We consider the pp, $\rm{{}^{7}Be}$ and pep data measured in the Borexino phase-I~\cite{Bellini:2011rx,Borexino:2011ufb,Borexino:2013zhu} and phase-II~\cite{Borexino:2017rsf,BOREXINO:2014pcl}, the $\rm{{}^{8}B}$ data~\cite{Borexino:2017uhp}, and the recent CNO data from Ref.~\cite{BOREXINO:2020aww}. The expected event rate is given by:
    \begin{equation}
        R_{\rm{pre}}^{i}=N_{e} \int d E_{\nu} \Phi^{i}\left(E_{\nu}\right)\left[P_{e e} \sigma_{e}\left(E_{\nu}\right)+\left(1-P_{es}-P_{ee}\right) \sigma_{\mu,\tau}\left(E_{\nu}\right)\right]\,,
        \label{eq.rate}
    \end{equation}
    where $i$ runs over solar neutrino sources $\rm{pp}$, $\rm{{}^{7}Be}$, $\rm{pep}$, $\rm{{}^{8}B}$ and $\rm{CNO}$, $\Phi^i$ is the corresponding neutrino flux from the standard solar model (B16-GS98-HZ)~\cite{Vinyoles:2016djt}, $P_{ee}$ and $P_{es}$ are evaluated by using the numerical method in Eq.~(\ref{eq:qppendix_nonad_pee_general}), $\sigma_{\alpha}$ is the cross-section given by
    \begin{align}
        \sigma_{\alpha}=\int dT_{e}\frac{d\sigma_\alpha}{dT_{e}}\eta(T_{e})\,,
    \end{align}
    where $\alpha=e,\mu,\tau$, $T_{e}$ is the recoil energy of electron, $\frac{d \sigma_{\alpha}}{dT_{e}}$ is the differential cross-section given by Eq.~(2.4) in Ref.~\cite{Chen:2021uuw}, and $\eta(T_{e})$ is the detection efficiency that is extracted from Fig.~2 in Ref.~\cite{Borexino:2017uhp} for $\rm{{}^{8}B}$ and set to be 100\% for other solar neutrino component~\cite{Khan:2019jvr}.
    The event rates observed by Borexino and our predicted event rates in the SM are shown in Table.~\ref{table.rate}. We can see  that our predictions in the SM are in good agreement with the measured results at Borexino.
    
\begin{table}[tbp]
	\centering
	\caption{The measured event rates at Borexino and our predicted event rates in the SM. The theoretical percentage uncertainties are given in the last column. }
	\label{table.rate}
	\begin{tabular}{|c|c|c|c|}
		\hline Source & Measurement (cpd/100 t) & SM prediction (cpd/100 t) & Percentage error \\
		\hline$\rm{pp}$ & $134 \pm 10_{-10}^{+6}$ & 136.1 & $1.2 \%$ \\
		\hline \multirow{2}*{$\rm{{}^{7}Be}$} & $46 \pm 1.5_{-1.6}^{+1.5}$ (phase I) & \multirow{2}*{47.4} & \multirow{2}*{$6.1 \%$} \\
		~ & $48.3 \pm 1.1_{-0.7}^{+0.4}$ (phase II) & ~ & ~ \\
		\hline \multirow{2}*{$\rm{pep}$} & $3.1 \pm 0.6 \pm 0.3$ (phase I) & \multirow{2}*{2.68} & \multirow{2}*{$1.3 \%$} \\
		~ & $2.43 \pm 0.36_{-0.22}^{+0.15}$ (phase II) & ~ & ~ \\
		\hline$\rm{{}^{8}B}$ & $0.223_{-0.016}^{+0.015} \pm 0.006$ & 0.233  & $12.0 \%$ \\
		\hline$\rm{CNO}$ & $7.2_{+3.0}^{-1.7}$ & 5.48 & $30.0 \%$ \\
		\hline
	\end{tabular}
\end{table}

		The KamLAND experiment is located in Kamioka mine, Gifu, Japan. It uses 1 kton of ultrapure liquid scintillator(LS) to monitor $\bar{\nu_e}$ flux from nuclear power reactors, and the flux-weighted average baseline is about 180 km. We use the same procedure as in Ref.~\cite{Liao:2017awz} to analyze the KamLAND data~\cite{KamLAND:2010fvi}, and obtain a preferred parameter region of $(\Delta m^{2}_{21},\sin^{2}\theta_{12})$ in the SM that agrees with Fig.~2 in \cite{Maltoni:2015kca}. In the following, we use the Borexino and KamLAND data to impose  constraints on the parameter space of super-light sterile neutrinos.

To evaluate the statistical significance of the new physics scenario, we define the $\chi^{2}$ function as follows:
\begin{equation}
\label{eq.chisq}
\chi^{2} = \chi^{2}_{\mathrm{Borexino}} + \chi^{2}_{\mathrm{KamLAND}}\,,
\end{equation}
where $\chi^{2}_{\mathrm{Borexino}}$ is taken from \cite{Khan:2019jvr}
\begin{equation}
\chi^{2}_{\mathrm{Borexino}} = \sum_{i} \left[\frac{R_{\rm{obs}}^{i}-R_{\rm{pre}}^{i}\left(1+\alpha^{i}\right)}{\sigma_{\rm{stat}}^{i}}\right]^{2}+\left(\frac{\alpha^{i}}{\sigma_{\rm{th}}^{i}}\right)^{2}\,,
\end{equation}
with $i$ running over the source listed in the first column in Table.~\ref{table.rate}. Here $R_{\rm{obs}}^{i}(\sigma_{\rm{obs}}^{i})$ are the central values (statistical uncertainties) of the $i^{\mathrm{th}}$ measurement given in Table.~\ref{table.rate}, $R_{\rm{pre}}^{i}$ is the predicted event rates calculated in Eq.~(\ref{eq.rate}), $\sigma_{\rm{stat}}^{i}$ is the experimental uncertainties given in the first column in Table.~\ref{table.rate}, and $\sigma^i_{\mathrm{th}}$ is the theoretical uncertainties that given in the last column in Table.~\ref{table.rate}.
The $\chi^{2}_{\mathrm{KamLAND}}$ in Eq.~(\ref{eq.chisq}) is given by
    \begin{equation}
    \chi^{2}_{\mathrm{KamLAND}} = \sum_{i} \left[\frac{N_{\rm{obs}}^{i}-N_{\rm{pre}}^{i}\left(1+\beta\right)}{\sigma_{\rm{error}}^{i}}\right]^{2}+\left(\frac{\beta}{\sigma_{\beta}}\right)^{2}\,,
\end{equation}
where $i$ represents the $i^{\rm{th}}$ energy bin, $N_{\rm{obs}}^i$ ($N_{\rm{pre}}^i$) is the event observed by KamLAND experiment (predicted in the 3+1 framework) in the $i^{\rm{th}}$ bin, $\sigma_{\rm{error}}^i$ is the experimental uncertainty  extracted from Fig.~1 in Ref.~\cite{KamLAND:2010fvi}, and $\sigma_{\beta}=0.043$ is the theoretical uncertainty of the reactor neutrino flux~\cite{KamLAND:2010fvi}.
In our analysis, we allow $\Delta m_{21}^{2}$ and $\theta_{12}$ to vary freely within the $3\sigma$ range of the global fit of neutrino oscillation data~\cite{2020}, while other oscillation parameters are fixed.

\subsection{Results}
We first focus on the case in which only $\theta_{01}$ is nonzero for the sterile mixing angles. The $95\% \ \rm{CL}$ bounds in the $\left(\sin^{2}2\theta_{01},\Delta m_{01}^{2}\right)$ plane are shown in Fig.~\ref{fig:contour_th01}. In particular, we show the bounds on the sterile parameter space for the low and high energy components of solar neutrinos separately. We can summarize the bounds from the Borexino data as follows:
\begin{figure}[tbp]
	\centering
	\includegraphics[width=0.7\textwidth]{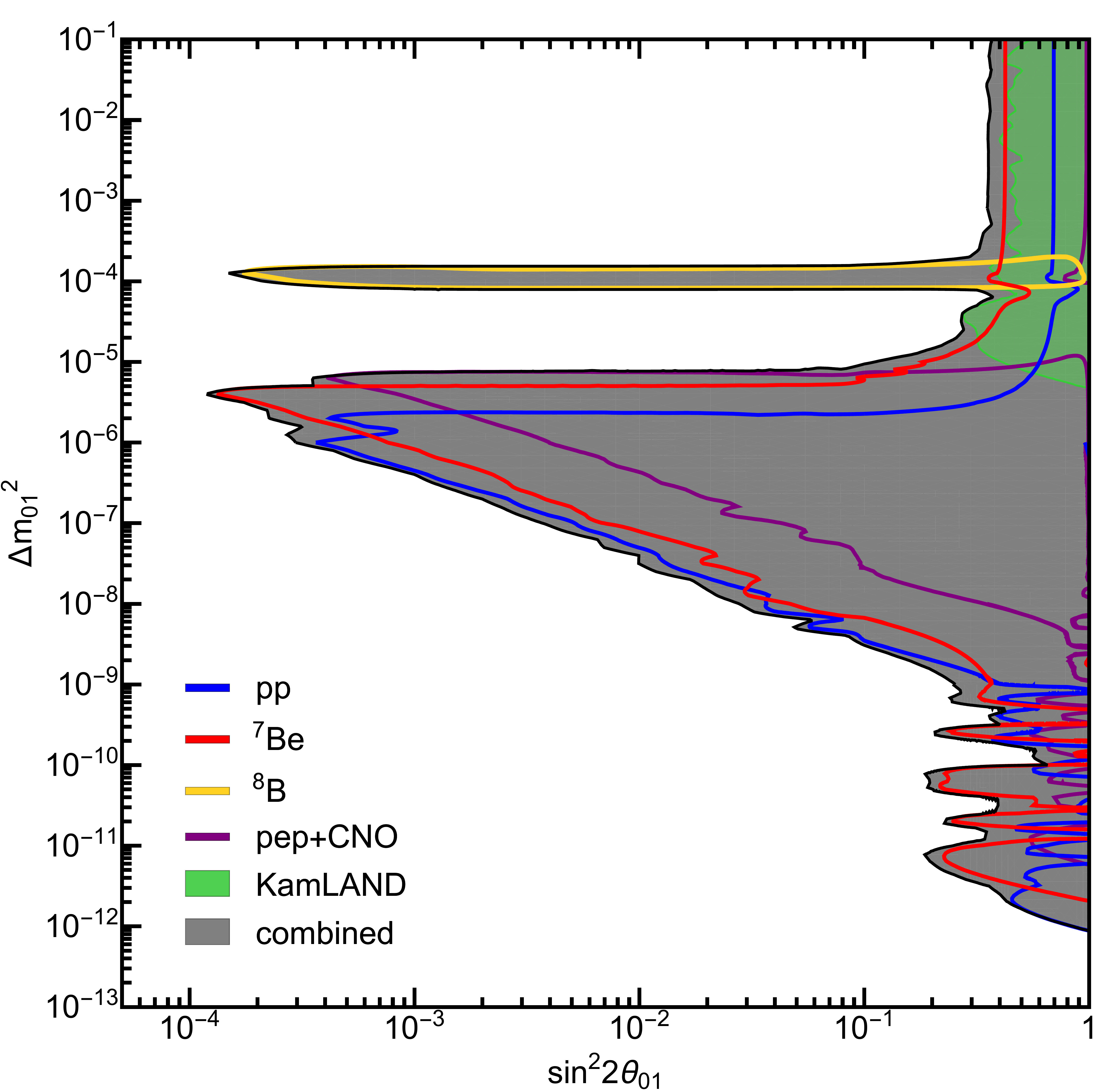}
	\caption{The 95 $\%$ CL bounds on parameters of sterile neutrino in the ($\sin^{2} 2\theta_{01}, \Delta m_{01}^{2}$) plane. Here we assume only $\theta_{01}$ is nonzero for the sterile mixing angles. The region enclosed by the blue, red, yellow and purple curves correspond to the exclusion regions obtained by using the $\rm{pp}$, $\rm{{}^{7}Be}$, $\rm{{}^{8}B}$ and $\rm{pep+CNO}$ data at Borexino, respectively. The green and gray shaded regions correspond to the exclusion regions obtained by the KamLAND data alone and combined data of all measurements at Borexino and KamLAND.}
	\label{fig:contour_th01}
\end{figure}
\begin{enumerate}[1.]
	\item For $\Delta m_{01}^{2} > 10^{-3} \ \textrm{eV}^{2}$, we can see that the bounds on the large mixing angle $\theta_{01}$ are set by the low energy neutrino $\left(\rm{pp}, \rm{{}^{7}Be}, \rm{pep+CNO}\right)$ data from Borexino. It can be understood from Eq.~(\ref{eq.big_th01_ad}), where we can see that at low energies, there is a large difference between $P_{ee}$ and $P_{ee}^{3\nu}$ if $\sin^{2}2\theta_{01}$ is large. By contrast, the constraints from the $\rm{{}^{8}B}$ data are very weak in this region. The reason can be explained by Eq.~(\ref{eq.big_th01_R_l_1}), where we get $P_{ee} \approx \left|U_{e2}\right|^{2} \approx 0.3$, and it approaches $P_{ee}^{3\nu}$ at high energies, as shown in Fig.~\ref{fig.posc_samples}(c). Therefore, $\rm{{}^{8}B}$ data does not impose significant constraints in this region.
	
	\item For $\Delta m_{01}^{2} \approx (1.1\sim2.2)\Delta m_{21}^{2}$, $\rm{{}^{8}B}$ data provides the best sensitivity for $\theta_{01}$, as shown in Fig.~\ref{fig:contour_th01}. The reason can be explained by the lower right panel of Fig.~\ref{fig.Pee_dip}. Since the measured energy range of the $\rm{{}^{8}B}$ neutrinos is from about 3 MeV to 12 MeV, and from Fig.~\ref{fig.Pee_dip}, we see that the resonance energy $E_{s1}$ for $1.1 \lesssim R_{\Delta} \lesssim 2.2$ also lies between 3 MeV and 12 MeV. Hence, the dip in $P_{ee}$ exactly locates within the measured energy range of $\rm{{}^{8}B}$ neutrinos, leading to a strong constraint from the $\rm{{}^{8}B}$ data in this region. From Fig.~\ref{fig:contour_th01}, we also see that the $\rm{{}^{8}B}$ data loses the sensitivity at $\sin^{2}2\theta_{01}\approx1$. This can be understood by Eq.~(\ref{eq.big_th01_R_g_1}), from which we get $P_{ee} \approx 0.35$ when $\sin^{2}2\theta_{01}\approx1$. Although $P_{ee}$ in this case differs from $P_{ee}^{3\nu}$ at $\sim$15\%, such a difference can be compensated by the large uncertainty $(\sim 12\%)$ in the $\rm{{}^{8}B}$ flux.
	
	\item For $10^{-9} \ \rm{eV^{2}} \lesssim \Delta m_{01}^{2} \lesssim 10^{-5} \ \rm{eV^{2}}$, the bounds are dominated by the low energy neutrino data. 
	From Fig.~\ref{fig:contour_th01}, one can see that the bounds set by $\rm{pp}$ $(\rm{{}^{7}Be})$ $[\rm{pep+CNO}]$ data become flat at about $2 \times 10^{-6}$ $(4 \times 10^{-6})$ $[7 \times 10^{-6}]$ $\rm{eV^{2}}$, which corresponds to $R_{\Delta}=$ 0.03 (0.07) [0.11]. This can be understood from the lower left panel of Fig.~\ref{fig.Pee_dip}. We can see from Fig.~\ref{fig.Pee_dip} that, for $R_{\Delta} = 0.03$ $(0.07)$ $[0.11]$, the position of the dip lies at 0.38 (0.86) [1.44] MeV, which corresponds to the peak energy of $\rm{pp}$ $(\rm{{}^{7}Be})$ $[\rm{pep}]$, respectively. However, the bounds can be weakened as $\Delta m_{01}^{2}$ or $\sin^22\theta_{01}$ becomes smaller, which is mainly caused by two factors: (i) as $\Delta m_{01}^{2}$ decreases, the position of the dip shifts to a lower energy, or even below from the energy region of solar neutrinos, so that $P_{ee}$ becomes closer to $P_{ee}^{3\nu}$; (ii) the hopping probability is also enhanced for smaller $\Delta m_{01}^{2}$ or $\sin^22\theta_{01}$, which results in a larger non-adiabatic transition between $\nu_{0}^{M}$ and $\nu_{1}^{M}$, and $P_{ee}$ gradually approaches $P_{ee}^{3\nu}$ as $\Delta m_{01}^{2}$ or $\sin^22\theta_{01}$ decreases; see Fig.~\ref{fig.Pee_vs_R}.
	On the contrary, $\rm{{}^{8}B}$ data is not sensitive to sterile neutrinos in this region. This is because $P_{ee}$ approaches $P_{ee}^{3\nu}$ as $E_{\nu}$ becomes large; see the blue zone of Fig.~\ref{fig.Pee_AD_vs_NonAD} for instance. In addition, from the lower left panel of Fig.~\ref{fig.Pee_dip} and Fig.~\ref{fig.posc_samples}(a), we can also see that $P_{ee}$ approaches $P_{ee}^{3\nu}$ at high energies. Therefore, the constraints from the $\rm{{}^{8}B}$ data in this region are very weak.
	
	\item For $\Delta m_{01}^{2} \lesssim 10^{-9} \ \textrm{eV}^{2}$, the bounds are mainly from the low energy neutrino data and are caused by the coherence of the quasi-degenerate states between $\nu_{0}$ and $\nu_{1}$. Since the amplitude of the coherent oscillation is proportional to $\sin2\theta_{01}$; see the third term in Eq.~(\ref{eq.pee_quasi_2f}), the bounds becomes weaker as $\theta_{01}$ decreases; see Fig.~{\ref{fig.Pee_coh_eng}}. At $\sin^{2}2\theta_{01}\simeq1$, the lowest bounds on the magnitude of $\Delta m_{01}^{2}$ from the $\rm{pp}$ data can reach $10^{-12} \ \rm{eV^{2}}$.~\footnote{Note that our lowest bound on the sterile mass-squared difference is about 10 times larger than the bound of Fig.6 in Ref.~\cite{deGouvea:2021ymm}, which is because the distance of Eq.~(\ref{eq.phase}) in our paper is less than the one of Eq.(IV.14) in Ref.~\cite{deGouvea:2021ymm} by a factor of 10.}
\end{enumerate}

	We also show the bounds on the sterile neutrino parameter space from the KamLAND data alone in Fig.~\ref{fig:contour_th01}, which mainly locates in the region with $\Delta m_{01}^{2} \gtrsim 5\times10^{-5} \ \textrm{eV}^{2}$. This can be understood from Eqs.~(\ref{eq.kam_02=0_h}) and (\ref{eq.kam_02=0_m}) in Appendix.~\ref{sec:kam}, since there is a large difference between $P_{ee}$ and $P_{ee}^{3\nu}$ if the sterile mixing angle $\theta_{01}$ is not small. However, for $\Delta m_{01}^{2} \ll 10^{-5} \ \textrm{eV}^{2}$, the constraints from KamLAND data is very weak. This is because if $\Delta m_{01}^{2}$ is very small, the oscillation length due to sterile neutrinos will be much larger than the baseline of KamLAND, and we can get $P_{ee} \approx P_{ee}^{3\nu}$ from Eq.~(\ref{eq.kam_02=0_l}). As a result, KamLAND data is not sensitive to sterile neutrinos in this region.
	We also obtain the exclusion regions of the combined data from all measurements at Borexino and KamLAND, which is shown as the gray shaded regions in Fig.~\ref{fig:contour_th01}. We can see that the combined bounds are dominated by the Borexino data. We find that the lowest combined bounds on $\sin^{2}2\theta_{01}$ can reach $1.3\times10^{-4}$ at $\Delta m_{01}^{2} = 4.0\times10^{-6} \ \rm{eV^{2}}$.


	\begin{figure}[tbp]
	\centering
	\includegraphics[width=0.7\textwidth]{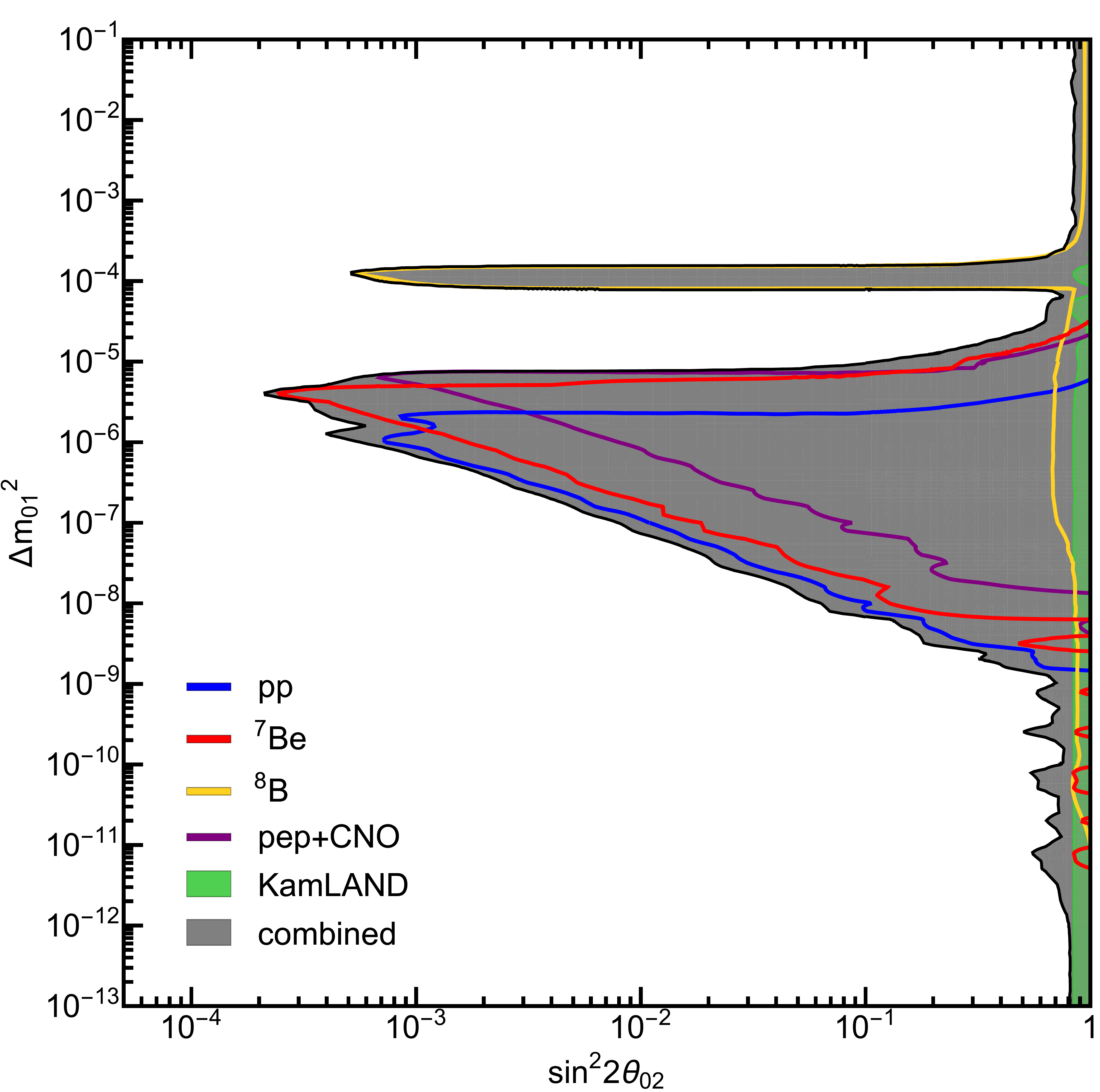}
	\caption{Same as in Fig.~\ref{fig:contour_th01}, except for $\theta_{02}$.}
	\label{fig:contour_th02}
\end{figure}
 
	In addition, we study the case in which $\nu_{s}$ consists of $\nu_{0}$ and $\nu_{2}$ in vacuum, i.e., only $\theta_{02}$ is nonzero for the sterile mixing angles. The bounds in the $\left(\sin^{2}2\theta_{02},\Delta m_{01}^{2}\right)$ plane are shown in Fig.~\ref{fig:contour_th02}. 
	From Fig.~\ref{fig:contour_th02}, we see that for the low energy neutrino data at Borexino, their sensitivity for $\Delta m_{01}^{2} > 10^{-3} \ \textrm{eV}^{2}$ is very weak. This can be understood from Eq.~(\ref{eq.big_th02_ad}). Since $\left|U_{e2}^{3\nu}\right|^{4} \approx 0.09$ is much smaller than $\left|U_{e1}^{3\nu}\right|^{4} \approx 0.5$, $P_{ee}$ is much closer to $P_{ee}^{3\nu}$ at low energies, resulting in the lost of the sensitivity for the low energy neutrino data in this case. In addition, the low energy neutrino data sets similar bounds for $10^{-9} \ \rm{eV^{2}} \lesssim \Delta m_{01}^{2} \lesssim 10^{-5} \ \rm{eV^{2}}$ as compared to the case with only $\theta_{01}$ being nonzero. However, the bounds in $\Delta m_{01}^{2} \ll 10^{-9} \ \rm{eV^{2}}$ are much weaker in this case. For the $\rm{{}^{8}B}$ data at Borexino, we find that it also sensitive to the low mass-squared difference region. The lowest bounds on $\Delta m_{01}^{2}$ from $\rm{{}^{8}B}$ data can reach $10^{-11} \ \textrm{eV}^{2}$. The reason for the enhanced sensitivity of $\rm{{}^{8}B}$ data is due to the fact that the sterile mixing angle $\theta_{02}$ has a large impact on $\left|U_{e2}\right|^{2}$ in this case, which leads to a large difference between $P_{ee}$ and $P_{ee}^{3\nu}$ at high energies for large $\theta_{02}$; see Eq.~(\ref{eq.big_th02_R_l_1}). By contrast, from Fig.~\ref{fig:contour_th01} we see that $\rm{{}^{8}B}$ data is not sensitive to the region of $\Delta m_{01}^{2} \ll 10^{-4} \ \rm{eV^{2}}$. This is because the sterile mixing angle $\theta_{01}$ mainly affects $\left|U_{e1}\right|^{2}$ in this case, and $P_{ee}$ at high energies always approaches $P_{ee}^{3\nu}$. In addition, the 95\% CL bounds set by the KamLAND data alone can be also seen from the green shadowed regions in Fig.~\ref{fig:contour_th02}. From Fig.~\ref{fig:contour_th02}, we see that it can set bounds for $\Delta m_{01}^{2} < 2.1 \Delta m_{21}^{2}$. This can be understood by Eq.~(\ref{eq.kam_01=0_l}), where we find there is a non-negligible difference between $P_{ee}$ and $P_{ee}^{3\nu}$ if $\sin^{2}2\theta_{02}$ is large. However, for $\Delta m_{01}^{2} \approx \Delta m_{21}^{2}$, the bounds from KamLAND data becomes weak, since we get $P_{ee}$ approaches $P_{ee}^{3\nu}$ from Eq.~(\ref{eq.kam_01=0_m}). For $\Delta m_{01}^{2} \gg \Delta m_{21}^{2}$, we find the difference between $P_{ee}$ and $P_{ee}^{3\nu}$ is small even if $\sin^{2}2\theta_{02}$ is large; see Eq.~(\ref{eq.kam_01=0_h}). Hence, KamLAND data is not sensitive to these regions.
From Fig.~\ref{fig:contour_th02}, we find that the strongest bound on $\sin^{2}2\theta_{02}$ can reach $2.2\times10^{-4}$ at $\Delta m_{01}^{2} = 4.0\times10^{-6} \ \rm{eV^{2}}$ from the combined data.

	\begin{figure}[tbp]
	\centering
	\includegraphics[width=0.7\textwidth]{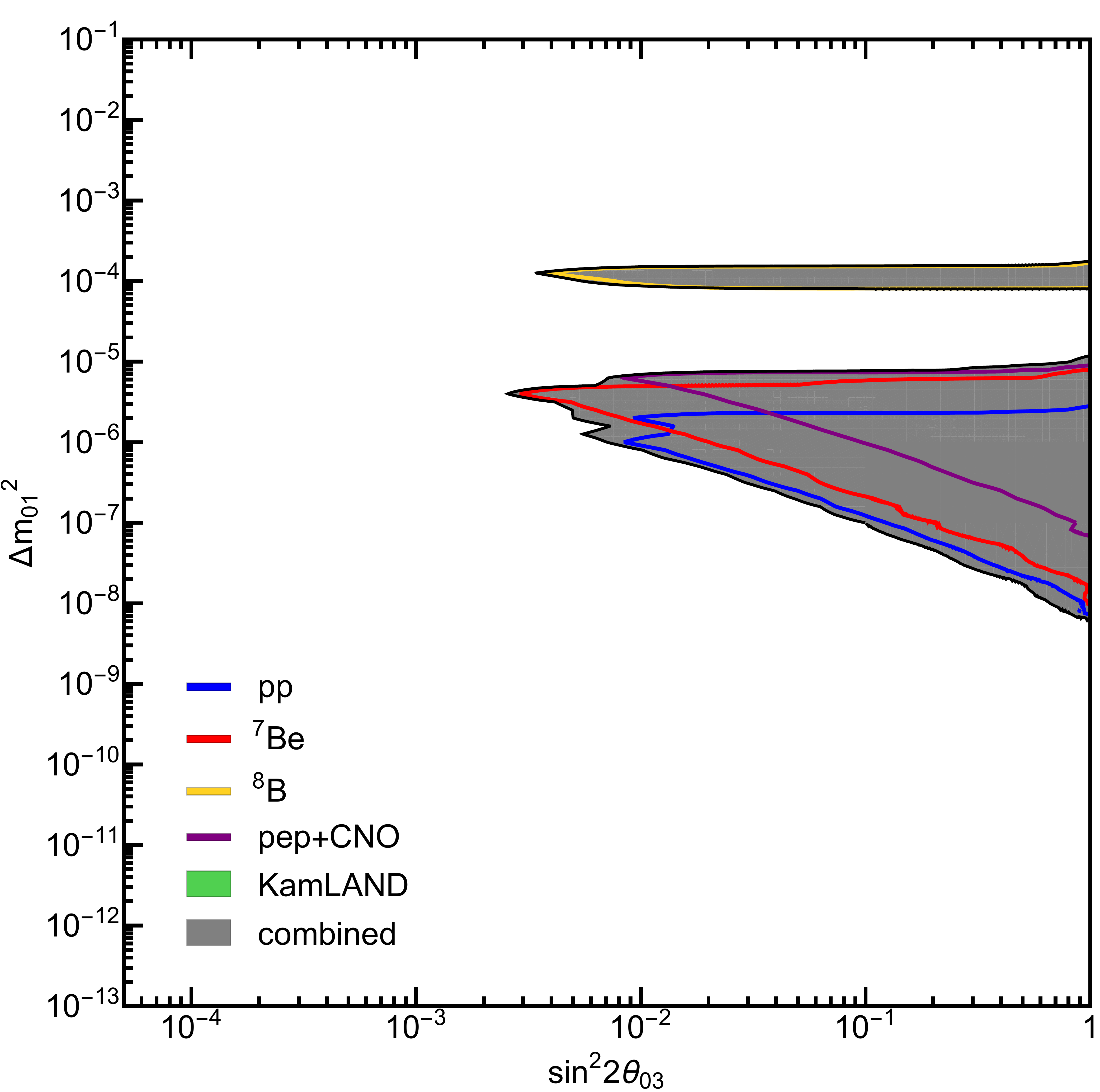}
	\caption{Same as in Fig.~\ref{fig:contour_th01}, except for $\theta_{03}$.}
	\label{fig:contour_th03}
\end{figure}

We also study the case in which $\nu_{s}$ consists of $\nu_{0}$ and $\nu_{3}$ in vacuum, i.e., only $\theta_{03}$ is nonzero for the sterile mixing angles. The bounds in the $\left(\sin^{2}2\theta_{03},\Delta m_{01}^{2}\right)$ plane are shown in Fig.~\ref{fig:contour_th03}. As we can see from Fig.~\ref{fig:contour_th03}, the low energy data at Borexino can set the bounds for $10^{-8} \ \rm{eV^{2}} \lesssim \Delta m_{01}^{2} \lesssim 10^{-5} \ \rm{eV^{2}}$, which are similar but much weaker than those in Fig.~\ref{fig:contour_th01}. We also see that the low energy neutrino data is not sensitive to the region of $\Delta m_{01}^{2} > 10^{-3} \ \rm{eV^{2}}$. It can be understood from Eq.~(\ref{eq.big_th03_ad}), where we find that $P_{ee}$ is approximately equal to $P_{ee}^{3\nu}$ since $\left|U_{e3}^{3\nu}\right|^{4}$ is negligible. For the ${}^{8}\rm{B}$ data at Borexino, it sets a similar but weaker bound for $\Delta m_{01}^{2} \approx (1.1\sim2.2)\Delta m_{21}^{2}$ as compared to the case with only $\theta_{01}$ being nonzero. This can be understood from Eq.~(\ref{eq.big_th03_R_g_1}), in which we see $P_{ee}$ at high energies is much smaller than $P_{ee}^{3\nu}$, leading to a strong sensitivity of ${}^{8}\rm{B}$ data for $\Delta m_{01}^{2} \approx (1.1\sim2.2)\Delta m_{21}^{2}$. However, from Fig.~\ref{fig:contour_th03}, we see that the bounds set by ${}^{8}\rm{B}$ data for other regions are very weak. In addition, the 95\% CL bounds from KamLAND data can not be found in Fig.~\ref{fig:contour_th03}. The reason is that the sterile mixing angle $\theta_{03}$ appears in $\left|U_{e0}\right|^{2}$ and $\left|U_{e3}\right|^{2}$ in this case, and since both $\left|U_{e0}\right|^{2}$ and $\left|U_{e3}\right|^{2}$ are suppressed by $\sin^{2}\theta_{13}$, the difference between $P_{ee}$ and $P_{ee}^{3\nu}$ becomes very small. Therefore, the constraints from KamLAND in this case are very weak.
From Fig.~\ref{fig:contour_th03}, we find that the combined bounds are also dominated by the Borexino data. The strongest combined bounds on $\sin^{2}2\theta_{03}$ can reach $2.8\times10^{-3}$ at $\Delta m_{01}^{2} = 4.0\times10^{-6} \ \rm{eV^{2}}$.

\section{Conclusion}
\label{sec:Con}
In this work, we studied the constraints on the parameter space of a super-light sterile neutrino imposed by the current Borexino and KamLAND data. The presence of a super-light neutrino can lead to a large modification to the survival probability of solar neutrinos.
We develop a numerical method to calculate the survival probability of solar neutrinos in the 3+1 framework by taking into account the non-adiabatic transitions and coherence effect between different mass eigenstates. 

To understand the effect of the super-light neutrino on the survival probability, we consider various scenarios with different combination of sterile neutrino parameters both numerically and analytically. 
We find that solar neutrinos can pass through two resonance points at high energies as they propagate through the Sun for $0<R_{\Delta} \lesssim 0.4 $, while there is only one level-crossing point at high energies for $R_{\Delta}>1$.
Also, for a small sterile mixing angle $\theta_{01}$, we find that there is a dip in the $P_{ee}$, and the position of the dip is determined by the first resonance energy $E_{s1}$, which can be predicted by an analytic equation. As $R_{\Delta}$ increases, the position of the dip shifts to a higher energy region. Due to the presence of the non-adiabatic transitions, the dip becomes weaker, and even disappears if two level crossing points occurr. 
In addition, we study the dependence of the survival probability on the sterile mixing angle, and find that the hopping probability can be effectively suppressed by increasing the sterile mixing angle. We also obtained an analytic equation of the survival probabilities for the quasi-degenerate scenarios by taking into account of both the non-adiabatic transitions and the coherence effect. The predictions from our analytic equation agree very well with the numerical results, and we find that the oscillations mainly occur in the low energy regions due to the coherence of quasi-degenerate states.  

We further set constraints on the parameter space of sterile neutrinos by using the latest Borexino and KamLAND data. In particular, we show the bounds on the sterile parameter space for the low and high energy components of solar neutrinos separately.
In the case with only $\theta_{01}$ being nonzero, we find that $\rm{{}^{8}B}$ data set the strongest bounds for $\Delta m_{01}^{2} \approx \left(1.1\sim2.2\right) \Delta m_{21}^{2}$, while the low energy neutrino data is more sensitive to the region with $10^{-9} \ \textrm{eV}^{2} \lesssim \Delta m_{01}^{2} \lesssim 10^{-5} \ \textrm{eV}^{2}$. As for the quasi-degenerate case, the bounds on $\Delta m_{01}^{2}$ from the $\rm{pp}$ data can reach $10^{-12} \ \rm{eV^{2}}$. We also find that the KamLAND data is only sensitive for $\Delta m_{01}^{2} > 10^{-5} \ \textrm{eV}^{2}$ in this case, and the combined bounds are dominated by the Borexino data.

We also study the case with only $\theta_{02}$ being nonzero. We find that the bounds from the low energy neutrino data are similar to the case with only $\theta_{01}$ being nonzero for $10^{-9} \ \rm{eV^{2}} \lesssim \Delta m_{01}^{2} \lesssim 10^{-5} \ \rm{eV^{2}}$, but becomes much weaker for $\Delta m_{01}^{2} \gg \Delta m_{21}^{2}$. For the $\rm{{}^{8}B}$ data at Borexino, we find that it is sensitive to both the low and high mass-squared difference regions, and the lowest bounds on $\Delta m_{01}^{2}$ from $\rm{{}^{8}B}$ data can reach $10^{-11} \ \textrm{eV}^{2}$. The KamLAND data can also set bounds for $\Delta m_{01}^{2} \lesssim 2.1 \Delta m_{21}^{2}$ if $\theta_{02}$ is nonzero. 
Furthermore, we study the case with only $\theta_{03}$ being nonzero. Compared to the case with only $\theta_{01}$ being nonzero, we find the low energy neutrino data set similar but much weaker bounds for $10^{-8} \ \textrm{eV}^{2} \lesssim \Delta m_{01}^{2} \lesssim 10^{-5} \ \textrm{eV}^{2}$. Also, the $\rm{{}^{8}B}$ data can also set a bound for $\Delta m_{01}^{2} \approx (1.1\sim2.2)\Delta m_{21}^{2}$. 

\acknowledgments
We thank P. C. de Holanda and A. Yu. Smirnov for helpful correspondence. 
J. Liao acknowledges the support from the National Natural Science Foundation of China 
under Grant No. 11905299, 
Guangdong Basic and Applied Basic Research Foundation under Grant No.~2020A1515011479, 
the Fundamental Research Funds for the Central Universities, and the Sun Yat-Sen University Science Foundation.
J. Ling acknowledges the support from National Key R\&D program of China under Grant NO. 2018YFA0404013, National Natural Science Foundation of China under Grant NO. 11775315, Key Lab of Particle \& Radiation Imaging, Ministry of Education. 
\newpage
\appendix

\section{The mixing matrix in the 3+1 framework}
\label{sec:Um}
In Appendix.~\ref{sec:Um}, we summarize the mixing matrix $U$ in the 3+1 framework. In general, $U$ in Eq.~(\ref{eq.U}) can be written as
\begin{align}
    U  & = R_{23}(\theta_{23})R_{13}(\theta_{13},\delta_{13})R_{12}(\theta_{12})R_{01}(\theta_{01},\delta_{01})R_{02}(\theta_{02},\delta_{02})R_{03}(\theta_{03},\delta_{03})\,.
\end{align}
For convenience, we set all phases to be zero in our work. If $\nu_{s}$ only mixes in $\nu_{0}$ and $\nu_{1}$ in vacuum, i.e., only $\theta_{01}$ is nonzero in the sterile mixing angles, then $U$ can be written as
\begin{equation}
	\label{eq.U_01}
	\begin{aligned}
		U    & = R_{23}(\theta_{23}) \cdot R_{13}(\theta_{13}) \cdot R_{12}(\theta_{12}) \cdot R_{01}(\theta_{01})\,, \\
		& = \left(\begin{array}{cccc}
			c_{01} & s_{01} & 0 & 0 \\
			-c_{13} c_{12} s_{01} & c_{13} c_{12} c_{01} & c_{13} s_{12} & s_{13} \\
			s_{01}(c_{23}s_{12}+c_{12} s_{23} s_{13}) & -c_{01}(c_{23}s_{12}+c_{12} s_{23} s_{13}) & c_{23} c_{12}-s_{23} s_{13} s_{12} & c_{13} s_{23} \\
			-s_{01}(s_{23} s_{12}-c_{23} c_{12} s_{13}) & c_{01}(s_{23} s_{12}-c_{23} c_{12} s_{13}) & -c_{12} s_{23}-c_{23} s_{13} s_{12} & c_{23} c_{13}
		\end{array}\right)\,.
	\end{aligned}
\end{equation}
If $\nu_{s}$ only mixes in $\nu_{0}$ and $\nu_{2}$ in vacuum, i.e., only $\theta_{02}$ is nonzero in the sterile mixing angles, then $U$ can be written as
\begin{equation}
	\label{eq.U_02}
	\begin{aligned}
		U    & = R_{23}(\theta_{23}) \cdot R_{13}(\theta_{13}) \cdot R_{12}(\theta_{12}) \cdot R_{02}(\theta_{02})\,, \\
		& = \left(\begin{array}{cccc}
			c_{02} & 0 & s_{02} & 0 \\
			-c_{13} s_{12} s_{02} & c_{13} c_{12} & c_{13} c_{02} s_{12} & s_{13} \\
			-s_{02}(c_{23}c_{12}-s_{23} s_{13} s_{12}) & -(c_{23}s_{12}+c_{12} s_{23} s_{13}) & c_{02}(c_{23} c_{12}-s_{23} s_{13} s_{12}) & c_{13} s_{23} \\
			s_{02}(c_{12} s_{23}-c_{23} s_{13} s_{12}) & s_{23} s_{12}-c_{23} c_{12} s_{13} & -c_{02}(c_{12} s_{23}+c_{23} s_{13} s_{12}) & c_{23} c_{13}
		\end{array}\right)\,.
	\end{aligned}
\end{equation}
If $\nu_{s}$ only mixes in $\nu_{0}$ and $\nu_{3}$ in vacuum, i.e., only $\theta_{03}$ is nonzero in the sterile mixing angles, then $U$ can be written as
\begin{equation}
	\label{eq.U_03}
	\begin{aligned}
		U    & = R_{23}(\theta_{23}) \cdot R_{13}(\theta_{13}) \cdot R_{12}(\theta_{12}) \cdot R_{03}(\theta_{03})\,, \\
		& = \left(\begin{array}{cccc}
			c_{03} & 0 & 0 & s_{03} \\
			-s_{13}s_{03} & c_{13} c_{12} & c_{13}s_{12} &  c_{03}s_{13} \\
			-c_{13}s_{23}s_{03} & -(c_{23}s_{12}+c_{12} s_{23} s_{13}) & c_{23} c_{12}-s_{23} s_{13} s_{12} & c_{13}c_{03}s_{23} \\
			-c_{13}c_{23}s_{03} & s_{23} s_{12}-c_{23} c_{12} s_{13} & -c_{12} s_{23}-c_{23} s_{13} s_{12} & c_{23}c_{13}c_{03}
		\end{array}\right)\,.
	\end{aligned}
\end{equation}
If both $\theta_{01}$ and $\theta_{02}$ are not equal to zero, i.e., $U = R_{23}(\theta_{23})R_{13}(\theta_{13})R_{12}(\theta_{12})R_{01}(\theta_{01})R_{02}(\theta_{02})$, the elements of $U$ is listed in Table.~\ref{table.U}. The effective mixing matrix in matter $U^{M}$ is similar to $U$ only by replacing the mixing angle $\theta_{ij}$ with the effective mixing angle $\theta_{ij}^{M}$. 
\begin{table}[htbp]
\centering
\caption{The elements of the mixing matrix in the 3+1 framework. Here, the mixing angle $\theta_{03}$ and all phases are assumed to be zero.}
\label{table.U}
\begin{tabular}{|c|c|c|}
\hline 
$\alpha$    &   $U_{\alpha i}$  &  -    \\
\cline{2-3}
\hline 
\multirow{4}*{$s$}  
&   $U_{s0}$    &   $c_{02}c_{01}$ \\
\cline{2-3}
&   $U_{s1}$    &   $s_{01}$ \\
\cline{2-3}
&   $U_{s2}$    &   $c_{01}s_{02}$ \\
\cline{2-3}
&   $U_{s3}$    &   0 \\
\hline 
\multirow{4}*{$e$}  
&   $U_{e0}$    &   $-c_{13}c_{12}c_{02}s_{01}-c_{13}s_{12}s_{02}$ \\
\cline{2-3}
&   $U_{e1}$    &   $c_{13}c_{12}c_{01}$ \\
\cline{2-3}
&   $U_{e2}$    &   $c_{13}c_{02}s_{12}-c_{13}c_{12}s_{02}s_{01}$ \\
\cline{2-3}
&   $U_{e3}$    &   $s_{13}$ \\
\hline 
\multirow{4}*{$\mu$}    
&   $U_{\mu0}$  &   $c_{02}s_{01}(c_{23}s_{12}+c_{12}s_{23}s_{13})-s_{02}(c_{23}c_{12}-s_{23}s_{13}s_{12})$ \\
\cline{2-3}
&   $U_{\mu1}$  &   $-c_{01}(c_{23}s_{12}+c_{12}s_{23}s_{12})$ \\
\cline{2-3}
&   $U_{\mu2}$  &   $s_{02}s_{01}(c_{23}s_{12}+c_{12}s_{23}s_{13})+c_{02}(c_{23}c_{12}-s_{23}s_{13}s_{12})$ \\
\cline{2-3}
&   $U_{\mu3}$  &   $c_{13}s_{23}$ \\
\hline
\multirow{4}*{$\tau$}   
&   $U_{\tau0}$ &   $s_{02}(c_{23}s_{13}s_{12}+c_{12}s_{23})-c_{02}s_{01}(s_{23}s_{12}-c_{23}c_{12}s_{13})$ \\
\cline{2-3}
&   $U_{\tau1}$ &   $c_{01}(s_{23}s_{12}-c_{23}c_{12}s_{13})$ \\
\cline{2-3}
&   $U_{\tau2}$ &   $-c_{02}(c_{23}s_{13}s_{12}+c_{12}s_{23})-s_{02}s_{01}(s_{23}s_{12}-c_{23}c_{12}s_{13})$ \\
\cline{2-3}
&   $U_{\tau3}$ &   $c_{23}c_{13}$ \\
\hline
\end{tabular}
\end{table}
\section{A numerical method of probabilities in the 3+1 framework}
\label{sec:Schrodinger}
    In Appendix.~\ref{sec:Schrodinger}, we present our numerical method used in section.~\ref{sec:prob}. We obtain the 3+1 probabilities by solving the Schrodinger evolution equations in the flavor basis. 
    In the 3+1 framework, the Schrodinger evolution equations in the flavor basis is given by
    \begin{equation}
        \label{Eq:appendixB_schrodinger}
        i\frac{d}{dx}(|\nu_{s}\rangle,|\nu_{e}\rangle,|\nu_{\mu}\rangle,|\nu_{\tau}\rangle)^{\mathsf{T}} = H_{f} (|\nu_{s}\rangle,|\nu_{e}\rangle,|\nu_{\mu}\rangle,|\nu_{\tau}\rangle)^{\mathsf{T}}\,,
    \end{equation}
    where the Hamiltonian $H_{f}$ is given by Eq.~(\ref{eq.Hf_4v}).
    Introducing $\psi_{e\alpha} \equiv \langle \nu_{e} | \nu_{\alpha} \rangle$ as amplitude of $\nu_{e} \rightarrow \nu_{\alpha}$, Eq.~(\ref{Eq:appendixB_schrodinger}) becomes
    \begin{equation}
        \label{Eq:appendixB_schrodinger_amp}
        i\frac{d}{dx}(\psi_{es},\psi_{ee},\psi_{e\mu},\psi_{e\tau})^{\mathsf{T}} = H_{f} (\psi_{es},\psi_{ee},\psi_{e\mu},\psi_{e\tau})^{\mathsf{T}}\,,
    \end{equation}
    In the plane wave approximation, we have 
    \begin{equation}
        \label{Eq:appendixB_a_1}
        \begin{aligned}
            \psi_{e \alpha}(x) = A_{\alpha} \cdot \exp(-i \delta_{\alpha} x) = A_{\alpha} \cos{\delta_{\alpha} x} -i A_{\alpha} \sin{\delta_{\alpha} x}\,,
        \end{aligned}
    \end{equation}
    \begin{equation}
        \label{Eq:appendixB_a_2}
         \frac{d}{dx}\left[\psi_{e\alpha}(x)\right] = - A_{\alpha} \delta_{\alpha} \sin{\delta_{\alpha} x} - i A_{\alpha} \delta_{\alpha} \cos{\delta_{\alpha} x}\,.
    \end{equation}
    In order to separate the real part from the imaginary part, we introduce $y_{2\mathrm{n}+1}$ and $y_{2\mathrm{n}}$ and make the following substitution
    \begin{equation}
        \label{Eq:appendixB_y}
        y_{\mathrm{2n}}(x)=A_{\alpha} \delta_{\alpha} \cos({\delta_{\alpha} x}) \quad , \quad y_{2\mathrm{n}+1}(x)= A_{\alpha} \delta_{\alpha} \sin({\delta_{\alpha} x}) \quad , \quad n = 0, 1, 2...\,.
    \end{equation}
    Then, we have
    \begin{equation}
        \label{Eq:appendixB_rela_1}
        \left\{  
            \begin{array}{lr}  
                \psi_{\mathrm{es}}(x)    = y_{0}(x) - i y_{1}(x)\,, \\
                \psi_{\mathrm{ee}}(x)    = y_{2}(x) - i y_{3}(x)\,, \\
                \psi_{\mathrm{e\mu}}(x)  = y_{4}(x) - i y_{5}(x)\,, \\
                \psi_{\mathrm{e\tau}}(x) = y_{6}(x) - i y_{7}(x)\,,
            \end{array}  
        \right.
        \quad \rm{and} \quad 
        \left\{  
            \begin{array}{lr}  
                \psi^{\prime}_{\mathrm{es}}(x)       =  y_{0}^{\prime}(x) - i y_{1}^{\prime}(x)\,, \\
                \psi^{\prime}_{\mathrm{ee}}(x)       =  y_{2}^{\prime}(x) - i y_{3}^{\prime}(x)\,, \\
                \psi^{\prime}_{\mathrm{e\mu}}(x)     =  y_{4}^{\prime}(x) - i y_{5}^{\prime}(x)\,, \\
                \psi^{\prime}_{\mathrm{e\tau}}(x)    =  y_{6}^{\prime}(x) - i y_{7}^{\prime}(x)\,.
            \end{array}  
        \right.
    \end{equation}
    By using Eq.~(\ref{Eq:appendixB_rela_1}), we can transform the $4\times4$ complex differential equation Eq.~(\ref{Eq:appendixB_schrodinger_amp}) into the following $8\times8$ real differential equation,
    \begin{equation}
        \label{Eq:appendixB_8_ODE}
        \begin{aligned}
            \left(\begin{array}{c} y'_{0}(x) \\ y'_{1}(x) \\ y'_{2}(x) \\ y'_{3}(x) \\ y'_{4}(x) \\ y'_{5}(x) \\ y'_{6}(x) \\ y'_{7}(x) \end{array}\right) = 
            \left(\begin{array}{cccccccc}
                H_{00}^{\Im} & -H_{00}^{\Re} & H_{01}^{\Im} & -H_{01}^{\Re} & H_{02}^{\Im} & -H_{02}^{\Re} & H_{03}^{\Im} & -H_{03}^{\Re} \\ 
                H_{00}^{\Re} &  H_{00}^{\Im} & H_{01}^{\Re} &  H_{01}^{\Im} & H_{02}^{\Re} &  H_{02}^{\Im} & H_{03}^{\Re} &  H_{03}^{\Im} \\ 
                H_{10}^{\Im} & -H_{10}^{\Re} & H_{11}^{\Im} & -H_{11}^{\Re} & H_{12}^{\Im} & -H_{12}^{\Re} & H_{13}^{\Im} & -H_{13}^{\Re} \\ 
                H_{10}^{\Re} &  H_{10}^{\Im} & H_{11}^{\Re} &  H_{11}^{\Im} & H_{12}^{\Re} &  H_{12}^{\Im} & H_{13}^{\Re} &  H_{13}^{\Im} \\ 
                H_{20}^{\Im} & -H_{20}^{\Re} & H_{21}^{\Im} & -H_{21}^{\Re} & H_{22}^{\Im} & -H_{22}^{\Re} & H_{23}^{\Im} & -H_{23}^{\Re} \\ 
                H_{20}^{\Re} &  H_{20}^{\Im} & H_{21}^{\Re} &  H_{21}^{\Im} & H_{22}^{\Re} &  H_{22}^{\Im} & H_{23}^{\Re} &  H_{23}^{\Im} \\ 
                H_{30}^{\Im} & -H_{30}^{\Re} & H_{31}^{\Im} & -H_{31}^{\Re} & H_{32}^{\Im} & -H_{32}^{\Re} & H_{33}^{\Im} & -H_{33}^{\Re} \\ 
                H_{30}^{\Re} &  H_{30}^{\Im} & H_{31}^{\Re} &  H_{31}^{\Im} & H_{32}^{\Re} &  H_{32}^{\Im} & H_{33}^{\Re} &  H_{33}^{\Im} \\ 
            \end{array}\right)
            \left(\begin{array}{c} y_{0}(x) \\ y_{1}(x) \\ y_{2}(x) \\ y_{3}(x) \\ y_{4}(x) \\ y_{5}(x) \\ y_{6}(x) \\ y_{7}(x) \end{array}\right)\,,
        \end{aligned}
    \end{equation}
    where $H_{\mathrm{ij}}^\Re$ and $H_{\mathrm{ij}}^\Im$ $(\mathrm{i,j}=0, 1, 2, 3)$ represent the real part and the imaginary part of the elements of $H_{f}$, respectively. By solving Eq.~(\ref{Eq:appendixB_8_ODE}), we can get $y_{2\mathrm{n}+1}$, $y_{2\mathrm{n}}$ and then $\psi_{e\alpha}^{SS}$ $(\alpha = s, e, \mu, \tau)$. Here, the superscript 'SS' denotes the position at the surface of the Sun. As a result, the 3+1 probabilities observed on the Earth can be written as
    \begin{equation}
        \label{eq:qppendix_nonad_pee_general}
            P_{e\alpha} = \left|\sum_{i=0}^{3}\sum_{\beta=s,e,\mu,\tau}U_{\alpha i}e^{-i\frac{\Delta m_{i1}^{2}}{2E_{\nu}}L_{0}}U_{i\beta}^{\dagger}\psi_{e\beta}^{\rm{SS}}\right|^{2}\,.
    \end{equation}
    Here, Eq.~(\ref{eq:qppendix_nonad_pee_general}) is the same as Eq.~(\ref{eq:pee_nonad}) after the substitution of Eq.~(\ref{eq.A_ei}).
\section{Derivation of resonance energy \texorpdfstring{$E_{s}$}{Lg}}
\label{sec:resonance_energy}
In Appendix.~\ref{sec:resonance_energy}, we give the derivation of the resonance energy $E_{s}$ in Eq.~(\ref{eq.s_resonance}).
From Eqs.~(\ref{eq.lamda_0})~(\ref{eq.lamda_1}) and~(\ref{eq.lamda_2}), when $\lambda_{0} = \lambda_{1}$ or $\lambda_{0} = \lambda_{2}$ is satisfied, we have
\begin{equation}
\begin{scriptsize}
\begin{aligned}
& \frac{\Delta m_{01}^{2}}{2E_{\nu}}\cos^{2}{\theta_{01}} = \frac{\Delta m_{21}^{2}}{4E_{\nu}} + \frac{2V_{NC}+\tilde{V}_{CC}}{2} \pm \sqrt{\left(\frac{\Delta m_{21}^{2}}{4E_{\nu}}\cos{2\theta_{12}}-\frac{\tilde{V}_{CC}}{2}\right)^{2}+\left(\frac{\Delta m_{21}^{2}}{4E_{\nu}}\sin{2\theta_{12}}\right)^{2}}\,.
\end{aligned}
\end{scriptsize}
\end{equation}
After squaring both left and right handed sides and combining terms, we can get the same quadratic function of $E_{\nu}$ for both $\lambda_{0} = \lambda_{1}$ and $\lambda_{0} = \lambda_{2}$, i.e.,
\begin{equation}
a E_{\nu}^{2} + b E_{\nu} + c = 0\,,
\end{equation}
where the coefficients $a$, $b$, $c$ are defined as follows:
\begin{equation}
\begin{aligned}
a & \equiv \frac{4V_{NC}\left(V_{NC}+\tilde{V}_{CC}\right)}{\Delta m_{21}^{2}}\,, \\
b & \equiv \left(2V_{NC}+\tilde{V}_{CC}\right)\left(1-2R_{\Delta}\cos{^{2}\theta_{01}}\right)+\tilde{V}_{CC}\cos{2\theta_{12}}\,, \\
c & \equiv \Delta m_{01}^{2}\cos{^{2}\theta_{01}}\left(R_{\Delta}\cos{^{2}\theta_{01}}-1\right) = \Delta m_{21}^{2}R_{\Delta}\cos{^{2}\theta_{01}}\left(R_{\Delta}\cos{^{2}\theta_{01}}-1\right)\,.
\end{aligned}
\end{equation}
Then, the discriminant $\Delta \equiv b^{2}-4ac$ is given by
\begin{equation}
\begin{footnotesize}
\begin{aligned}
\Delta 
& = \left(\tilde{V}_{CC}+2V_{NC}\right)^{2} \left[\left(1-2R_{\Delta}\cos{^{2}\theta_{01}}+\xi\cos{2\theta_{12}}\right)^{2} - 4\left(\xi^{2}-1\right)R_{\Delta}\cos{^{2}\theta_{01}}\left(1-R_{\Delta}\cos{^{2}\theta_{01}}\right)\right]\,,
\label{eq.delta}
\end{aligned}
\end{footnotesize}
\end{equation}
where $\xi\equiv V_{CC}\cos{^{2}\theta_{13}}/(V_{CC}\cos{^{2}\theta_{13}}+2V_{NC})$. Using the extract root formula,
\begin{equation}
\frac{-b\pm\sqrt{\Delta}}{2a}=\frac{\left(-b\pm\sqrt{\Delta}\right)\left(-b\mp\sqrt{\Delta}\right)}{2a\left(-b\mp\sqrt{\Delta}\right)}=\frac{2c}{-b\mp\sqrt{\Delta}}\,,
\end{equation}
we can obtain the resonance energy $E_{s}$ as
\begin{equation}
\begin{footnotesize}
\begin{aligned}
& E_{s} = \frac{2\Delta m_{21}^{2}\cos{^{2}\theta_{01}}}{\tilde{V}_{CC}+2V_{NC}} \times \\
& \frac{R_{\Delta}(1-R_{\Delta} \cos{^{2}\theta_{01}})}{(1-2R_{\Delta} \cos{^{2}\theta_{01}} + \xi \cos{2\theta_{12}})\mp\sqrt{(1-2R_{\Delta} \cos{^{2}\theta_{01}} + \xi \cos{2\theta_{12}})^{2}-4(\xi^{2}-1)R_{\Delta} \cos{^{2}\theta_{01}}(1-R_{\Delta} \cos{^{2}\theta_{01}})}}\,.
\end{aligned}
\end{footnotesize}
\end{equation}
\section{Derivation of the analytic formulas of probabilities in the quasi-degenerate case}
\label{sec:s_matrix}
In Appendix.~\ref{sec:s_matrix}, we give the derivation of Eqs.~(\ref{eq.pee_quasi_pee}) and~(\ref{eq.pee_quasi_pes}) in section.~\ref{sec:degenerate}. Here, We emphasize that the analytic formulas we obtain has taken the non-adiabatic transitions and the coherence effect into account. We have checked its consistency by comparing with the numerical results in Fig.~\ref{fig.Pee_coh_eng}.
The neutrino oscillation probabilities can be given by
\begin{equation}
P_{\alpha \beta} = \left|S_{\beta \alpha}\left(t,t_{0}\right)\right|^{2}\,,
\end{equation}
where $S_{\beta \alpha}\left(t,t_{0}\right)$ is the evolution matrix, i.e.,
\begin{equation}
\begin{aligned}
|\nu_{\alpha}(t)\rangle = S(t,t_{0})|\nu_{\alpha}(t_{0})\rangle\,,
\end{aligned}
\end{equation}
with $|\nu_{\alpha}\rangle \equiv (|\nu_{s}\rangle,|\nu_{e}\rangle,|\nu_{\mu}\rangle,|\nu_{\tau}\rangle)^{\mathsf{T}}$.
Since the solar neutrino evolution can be described by Eq.~(\ref{eq.Hf_4v}), $S\left(x,x_{0}\right)$ can be written as
\begin{equation}
S = e^{-\frac{\textrm{i}}{\hbar}\int_{x_{0}}^{x} H_{f}(x')dx'}\,,
\end{equation}
In the adiabatic case, $S\left(x,x_{0}\right)$ can be obtained by diagonalizing the Hamiltonian in Eq.~(\ref{eq.Hf_4v}), 
\begin{equation}
S_{\beta \alpha}(x,x_{0}) = \sum_{i=0}^{3} U_{\beta i}U_{\alpha i}^{M_{0}} e^{-\textrm{i}E_{i}(x-x_{0})}
\end{equation}
where $E_{i}$ represents the eigenvalues of Eq.~(\ref{eq.Hf_4v}). In the center of the Sun, $|\nu_{\alpha}(0)\rangle = (0,1,0,0)$, therefore
\begin{equation}
\begin{aligned}
P_{ee} =  \left| \langle\nu_{e}| S |\nu_{\alpha}(0)\rangle \right|^{2}=\left|S_{e e}\left(t,0\right)\right|^{2} =\sum_{i=0}^{3} \left|U_{e i}\right|^{2}\left|U_{e i}^{M_{0}} \right|^{2}\,.
\end{aligned}
\end{equation}
In the non-adiabatic case, the solar neutrino evolution becomes complicated. However, for the quasi-degenerate case, i.e., $\Delta m_{01}^{2} \ll 10^{-9} \ \textrm{eV}^{2}$, we can also obtain the probabilities under some approximations. 
We consider that $\nu_{s}$ only mixes in $\nu_{0}$ and $\nu_{1}$ in vacuum with $\theta_{01}$.
Since $\nu_{3}$ decouples from other mass eigenstates in the Sun, we first rotate the Hamiltonian in the basis with $|\tilde{\nu_{\alpha}}\rangle = \tilde{U}^{\dagger} |\nu_{\alpha}\rangle$, i.e., 
\begin{equation}
\begin{aligned}
P_{ee}  
&= \left| \langle\nu_{e}| \tilde{U} \tilde{S'} (\tilde{U}^{M_{0}})^{\dagger} |\nu_{\alpha}(0)\rangle \right|^{2} \\
&= \left| \langle\nu_{e}| R_{23}R_{13}\tilde{S'} \left(R_{13}^{M_{0}}\right)^{\dagger}\left(R_{23}^{M_{0}}\right)^{\dagger} |\nu_{\alpha}(0)\rangle \right|^{2}\,,
\end{aligned}
\end{equation}
where $\tilde{S'}$ can be written as
\begin{equation}
\tilde{S'} = \left(
\begin{array}{ccc|c} 
\ \ \ \ \  &  \ \ \ \ \                & \ \ \ \ \  &     \\
\ \ \ \ \  &  (\tilde{S'})_{3\times3}  & \ \ \ \ \  &     \\
\ \ \ \ \  &  \ \ \ \ \                & \ \ \ \ \  &     \\
\hline              
\multicolumn{3}{c|}{ } & \tilde{\gamma_{3}}
\end{array}
\right)\,.
\end{equation}
Here, $(\tilde{S'})_{3\times3}$ is determined by the Hamiltonian in Eq.~(\ref{eq.H_new}), and $\tilde{\gamma_{3}} \approx \exp(-i\frac{\Delta m_{31}^{2}}{2E_{\nu}}x)$. For $\Delta m_{01}^{2} \ll 10^{-9} \ \textrm{eV}^{2}$, the off-diagonal terms in the first row of $H_{3\times3}$ are suppressed by $\Delta m_{01}^{2}$. To further simplify the evolution matrix $\tilde{S'}$, we rotate the basis with $R_{12}^{M}$, then
	\begin{align}
	P_{ee}  &= \left| \langle\nu_{e}| R_{23}R_{13}R_{12} \tilde{S''} \left(R_{12}^{M_{0}}\right)^{\dagger}\left(R_{13}^{M_{0}}\right)^{\dagger}\left(R_{23}^{M_{0}}\right)^{\dagger} |\nu_{\alpha}(0)\rangle \right|^{2} \\ 
	&\approx \left| \langle\nu_{e}| U^{' 3\nu} \tilde{S''} \left(U^{'M_{0}, 3\nu}\right)^{\dagger} |\nu_{\alpha}(0)\rangle \right|^{2} \\
	&= \left|\sum_{i,j=0}^{3}\left(U^{' 3\nu}\right)_{ei}\left(U^{'M_{0}, 3\nu}\right)_{ej} \left(\tilde{S''}\right)_{ij}\right|^{2} \label{eq.ij0} \\
	&= \left|\sum_{i,j=1}^{3}\left(U^{' 3\nu}\right)_{ei}\left(U^{'M_{0}, 3\nu}\right)_{ej} \left(\tilde{S''}\right)_{ij}\right|^{2} \label{eq.ij1}\,, 
	\end{align}
where Eq.~(\ref{eq.ij0}) turns into Eq.~(\ref{eq.ij1}) since both $\left(U^{' 3\nu}\right)_{e0}$ and $\left(U^{'M_{0}, 3\nu}\right)_{e0}$ are equal to zero. Here, $U^{' 3\nu}$ can be written as
\begin{equation}
\left(
\begin{array}{c|ccc} 
\ 1 \ \ & \ \ & \ \ & \ \ \\
\hline              
\  & \ \ \ \ \  &  \ \ \ \ \                & \ \ \ \ \    \\
\  & \ \ \ \ \  &  U^{3\nu} & \ \ \ \ \      \\
\  & \ \ \ \ \  &  \ \ \ \ \                & \ \ \ \ \    \\
\end{array}
\right)\,.
\end{equation}
$U^{'M_{0}, 3\nu}$ also has a similar form by replacing the sub-matrix $U^{3\nu}$ with $U^{M_{0}, 3\nu}$. Since $\left(\tilde{S''}\right)_{ij}$ $(i,j=1,2,3 \ {\rm and} \ i\neq j)$ are the terms that can be averaged out due to the large distance between the Sun and the Earth, then only the elements $\left(\tilde{S''}\right)_{22}$, $\left(\tilde{S''}\right)_{33}$ and the sub-matrix $\left(\tilde{S''}\right)_{2\times2}$ remain, so that $\tilde{S''}$ can be written as
\begin{equation}
\tilde{S''} = \left(
\begin{array}{cc|cc} 
\  &  \ \ \                   & \ \ \   &  \ \ \    \\
\ &  (\tilde{S''})_{2\times2}  & \ \ \   &  \ \ \    \\
\hline  
& & \tilde{\gamma_{2}} & \\
& &  \ \ \  & \tilde{\gamma_{3}}
\end{array}
\right)\,,
\end{equation}
where $\tilde{\gamma_{2}} \approx \exp(-i\lambda_{2}^{\rm{LMA}}x)$. Then, we have
\begin{equation}
P_{ee}  = \left|\sum_{i=1}^{3}\left(U^{' 3\nu}\right)_{ei}\left(U^{'M_{0}, 3\nu}\right)_{ei} \left(\tilde{S''}\right)_{ii}\right|^{2}\,, 
\end{equation}
Further, under the assumptions above, the mixing matrices have the following relationship
\begin{align}
&\left|\left(U^{'3\nu}\right)_{e1}\right|^{2} = 1-\left|U_{e2}\right|^{2}-\left|U_{e3}\right|^{2}\,, \label{eq.U1}\\
&\left(U^{' 3\nu}\right)_{e2} = U_{e2} \quad , \quad \left(U^{' 3\nu}\right)_{e3} = U_{e3}\,, \label{eq.U23}\\
&\left|\left(U^{'M_{0}, 3\nu}\right)_{e1}\right|^{2} = 1-\left|U^{M_{0}}_{e2}\right|^{2}-\left|U^{M_{0}}_{e3}\right|^{2}\,, \label{eq.U1m}\\
&\left(U^{'M_{0}, 3\nu}\right)_{e2} = U^{M_{0}}_{e2} \quad , \quad \left(U^{'M_{0}, 3\nu}\right)_{e3} = U^{M_{0}}_{e3}\,. \label{eq.U23m}
\end{align}
Therefore, we obtain
\begin{align}
    P_{ee}      =& \left(1-\left|U_{e2}\right|^{2}-\left|U_{e3}\right|^{2}\right)\left(1-\left|U_{e2}^{M_{0}}\right|^{2}-\left|U_{e3}^{M_{0}}\right|^{2}\right) P_{ee}^{2f} \notag \\
                 & + \left|U_{e2}\right|^{2}\left|U_{e2}^{M_{0}}\right|^{2} + \left|U_{e3}\right|^{2}\left|U_{e3}^{M_{0}}\right|^{2}\,, \\
    P_{es}      =& \left(1-\left|U_{e2}^{M_{0}}\right|^{2}-\left|U_{e3}^{M_{0}}\right|^{2}\right) \left(1-P_{ee}^{2f}\right)\,, 
\end{align}
where $P_{ee}^{2f}$ is given by
\begin{equation}
\begin{aligned}
P_{ee}^{2f} &= \left| \tilde{S''}_{11}\right|^{2} = \left| \langle\nu_{e}| \left(\tilde{S''}\right)_{2\times2} |\nu_{\alpha}(0)\rangle \right|^{2}\,.
\end{aligned}
\end{equation}
Here, $(\tilde{S''})_{2\times2}$ satisfies the Schrodinger equation in the $2\nu$ scheme,
\begin{equation}
\label{eq.hf_pee2f}
i\frac{d}{dt}
\left(\begin{array}{c} |\nu_{s}(x)\rangle \\ |\nu_{1}^{''}(x)\rangle \end{array}\right)
= \left[R_{01}  
\left(\begin{array}{cc} \frac{\Delta m_{01}^{2}}{2E_{\nu}} & 0 \\ 0 & 0 \end{array}\right)
R_{01}^{\dagger} +
\left(\begin{array}{cc}  0 & 0 \\ 0 & V_{CC}+V_{NC} \end{array}\right) \right]
\left(\begin{array}{c} |\nu_{s}(x)\rangle \\ |\nu_{1}^{''}(x)\rangle \end{array}\right)\,,
\end{equation}
where $|\nu_{1}^{''}\rangle = \left(U^{' 3\nu}\right)^{\dagger}_{e1} |\nu_{e}\rangle$. We can see from Eq.~(\ref{eq.hf_pee2f}) that $P_{ee}^{2f}$ can be considered as the survival probability in the case of two neutrino mixing. Such that $P_{ee}^{2f}$ can follow the expression of the survival probability in the $2\nu$ scheme~\cite{deGouvea:2000pqg},
\begin{align}
    P_{ee}^{2f} =& \sin^{2}\theta_{01} + \cos{2\theta_{01}}\left[P_{c}\sin^{2}\theta_{01}^{M_{0}}+\left(1-P_{c}\right)\cos^{2}\theta_{01}^{M_{0}}\right] \notag \\
                &-\sqrt{P_{c}\left(1-P_{c}\right)}\cos{2\theta_{01}^{M_{0}}}\sin{2\theta_{01}}\cos{\left(2.54\frac{\Delta m_{01}^{2}}{E_{\nu}}L_{0}+\phi\right)}\,.
\end{align}

If $\nu_{s}$ only mixes in $\nu_{0}$ and $\nu_{2}$ in vacuum with the mixing angle $\theta_{02}$, for $\left|\Delta m_{02}^{2}\right|^{2} \ll 10^{-9} \ \rm{eV^{2}}$, we also obtain the analytic formulas of oscillation probabilities. In this case, $P_{ee}$ is the same as Eq.~(\ref{eq.ij1}). But in this case,  $\tilde{S''}$ becomes
\begin{equation}
\tilde{S''} = \left(
\begin{array}{cccc} 
  \tilde{S''_{00}}   & \ & \tilde{S''_{02}} & \\
& \tilde{\gamma_{1}} & \ & \\
  \tilde{S''_{20}}   & \ & \tilde{S''_{22}} \ & \\
& \ & \ & \tilde{\gamma_{3}}
\end{array}
\right)\,.
\end{equation}
The relationship in Eqs.~(\ref{eq.U1})~(\ref{eq.U23})~(\ref{eq.U1m}) and~(\ref{eq.U23m}) become
\begin{align}
&\left|\left(U^{'3\nu}\right)_{e2}\right|^{2} = 1-\left|U_{e1}\right|^{2}-\left|U_{e3}\right|^{2}\,, \label{eq.U1_02}\\
&\left(U^{' 3\nu}\right)_{e1} = U_{e1} \quad , \quad \left(U^{' 3\nu}\right)_{e3} = U_{e3}\,, \label{eq.U23_02}\\
&\left|\left(U^{'M_{0}, 3\nu}\right)_{e2}\right|^{2} = 1-\left|U^{M_{0}}_{e1}\right|^{2}-\left|U^{M_{0}}_{e3}\right|^{2}\,, \label{eq.U1m_02}\\
&\left(U^{'M_{0}, 3\nu}\right)_{e1} = U^{M_{0}}_{e1} \quad , \quad \left(U^{'M_{0}, 3\nu}\right)_{e3} = U^{M_{0}}_{e3}\,. \label{eq.U23m_02}
\end{align}
By exchanging indicators $1\leftrightarrow2$, we obtain
\begin{align}
    P_{ee}      =& \left(1-\left|U_{e1}\right|^{2}-\left|U_{e3}\right|^{2}\right)\left(1-\left|U_{e1}^{M_{0}}\right|^{2}-\left|U_{e3}^{M_{0}}\right|^{2}\right) P_{ee}^{2f} \notag \\
                 & + \left|U_{e1}\right|^{2}\left|U_{e1}^{M_{0}}\right|^{2} + \left|U_{e3}\right|^{2}\left|U_{e3}^{M_{0}}\right|^{2}\,, \\
    P_{es}      =& \left(1-\left|U_{e1}^{M_{0}}\right|^{2}-\left|U_{e3}^{M_{0}}\right|^{2}\right) \left(1-P_{ee}^{2f}\right)\,, 
\end{align}
where
\begin{align}
    P_{ee}^{2f} =& \sin^{2}\theta_{02} + \cos{2\theta_{02}}\left[P_{c}\sin^{2}\theta_{02}^{M_{0}}+\left(1-P_{c}\right)\cos^{2}\theta_{02}^{M_{0}}\right] \notag \\
                &-\sqrt{P_{c}\left(1-P_{c}\right)}\cos{2\theta_{02}^{M_{0}}}\sin{2\theta_{02}}\cos{\left(2.54\frac{\Delta m_{02}^{2}}{E_{\nu}}L_{0}+\phi\right)}\,.
\end{align}

\section{ Oscillation probabilities in the 3+1 framework at reactor neutrino experiments}
\label{sec:kam}
%
For reactor neutrinos observed by KamLAND, the matter effect of the Earth is negligible that we can simply treat it as the oscillation in vacuum. The survival probability of $\bar{\nu}_{e}$ can be written as
\begin{equation}
\begin{aligned}
\label{eq.kam_gen}
P_{ee} =& 1 - 4\left( \left|U_{e0}\right|^{2}\left|U_{e1}\right|^{2}\sin^{2}\Delta_{01} + \left|U_{e0}\right|^{2}\left|U_{e2}\right|^{2}\sin^{2}\Delta_{02} +   
\left|U_{e0}\right|^{2}\left|U_{e3}\right|^{2}\sin^{2}\Delta_{03} \right.\\
&\left. + \left|U_{e1}\right|^{2}\left|U_{e2}\right|^{2}\sin^{2}\Delta_{21} + \left|U_{e1}\right|^{2}\left|U_{e3}\right|^{2}\sin^{2}\Delta_{31} + \left|U_{e2}\right|^{2}\left|U_{e3}\right|^{2}\sin^{2}\Delta_{32}^{2} \right)\,,
\end{aligned}
\end{equation}
where $\Delta_{ij} \equiv \frac{\Delta m_{ij}^{2}}{4E_{\nu}}L$. For simplicity, we assume $\theta_{03}=0$ and consider $|\Delta_{03}|\approx|\Delta_{31}|\approx|\Delta_{32}|\approx\Delta_{3i}$, then Eq.~(\ref{eq.kam_gen}) becomes
\begin{equation}
\label{eq.kam_03=0}
\begin{aligned}
P_{ee} =& 1 - 4\left[ \left|U_{e0}\right|^{2}\left|U_{e1}\right|^{2}\sin^{2}\Delta_{01} + \left|U_{e0}\right|^{2}\left|U_{e2}\right|^{2}\sin^{2}\Delta_{02} + \left|U_{e1}\right|^{2}\left|U_{e2}\right|^{2}\sin^{2}\Delta_{21} \right. \\
&\left.+ \left|U_{e3}\right|^{2}\left(1-\left|U_{e3}\right|^{2}\right)\sin^{2}\Delta_{3i}\right]\,.
\end{aligned}
\end{equation}
	In the SM case, 
$P_{ee}$ is given by
	\begin{equation}
	\label{eq.kam_sm}
	\begin{aligned}
	P_{ee}^{3\nu} 
	=& 1 - 4\left(\left|U_{e1}\right|^{2}\left|U_{e2}\right|^{2}\sin^{2}\Delta_{21}+\left|U_{e1}\right|^{2}\left|U_{e3}\right|^{2}\sin^{2}\Delta_{13}+\left|U_{e2}\right|^{2}\left|U_{e3}\right|^{2}\sin^{2}\Delta_{32} \right) \\
	=& 1 - 4\left|U_{e1}^{3\nu}\right|^{2}\left|U_{e2}^{3\nu}\right|^{2}\sin^{2}\Delta_{21}- 4\left|U_{e3}^{3\nu}\right|^{2}\left(1-\left|U_{e3}^{3\nu}\right|^{2}\right)\sin^{2}\Delta_{3i} \,.
	\end{aligned}
	\end{equation}
	If only $\theta_{01}$ is nonzero for the sterile mixing angles, we have $\left|U_{e0}\right|^{2}=s^{2}_{01}\left|U_{e1}^{3\nu}\right|^{2}=s^{2}_{01}c^{2}_{12}c^{2}_{13}$, $\left|U_{e1}\right|^{2}=c^{2}_{01}\left|U_{e1}^{3\nu}\right|^{2}=c^{2}_{01}c^{2}_{12}c^{2}_{13}$, $\left|U_{e2}\right|^{2}=\left|U_{e2}^{3\nu}\right|^{2}=s^{2}_{12}c^{2}_{13}$ and $\left|U_{e3}\right|^{2}=\left|U_{e3}^{3\nu}\right|^{2}=s^{2}_{13}$; see Eq.~(\ref{eq.U_01}) in Appendix.~\ref{sec:Um}. Then,
	
	\begin{enumerate}
		\item $\Delta m_{01}^{2} \gg \Delta m_{21}^{2}$.
		In this case, the phase $\Delta_{01}$, $\Delta_{02}$ of the sine in Eq.~(\ref{eq.kam_03=0}) is very large due to the long baseline in KamLAND. It has a variation much larger than $2\pi$ in the energy resolution interval of the detector. So, the survival probability $P_{ee}$ is given by
		\begin{equation}
		\label{eq.kam_02=0_h}
		\begin{aligned}
		P_{ee}  
		\approx & 1 - 2\left( \left|U_{e0}\right|^{2}\left|U_{e1}\right|^{2} + \left|U_{e0}\right|^{2}\left|U_{e2}\right|^{2}\right) - 4\left(\left|U_{e1}\right|^{2}\left|U_{e2}\right|^{2}\sin^{2}\Delta_{21} \right) \\
		& - 4 \left|U_{e3}\right|^{2}\left(1-\left|U_{e3}\right|^{2}\right)\sin^{2}\Delta_{3i} \\
		= & P_{ee}^{3\nu} - 2\left|U_{e0}\right|^{2}\left(\left|U_{e1}\right|^{2} + \left|U_{e2}\right|^{2}\right) + 4\left(\left|U_{e0}\right|^{2}\left|U_{e2}\right|^{2}\sin^{2}\Delta_{21} \right)
		\end{aligned}
		\end{equation}
		
		\item $\Delta m_{01}^{2} \sim \Delta m_{21}^{2}$.
		In this case, Eq.~(\ref{eq.kam_03=0}) becomes
		\begin{equation}
		\label{eq.kam_02=0_m}
		\begin{aligned}
		P_{ee}  
		\approx & 1 - 4\left( \left|U_{e0}\right|^{2}\left|U_{e1}\right|^{2}\sin^{2}\Delta_{01} + \left|U_{e0}\right|^{2}\left|U_{e2}\right|^{2}\sin^{2}\Delta_{02} + \left|U_{e1}\right|^{2}\left|U_{e2}\right|^{2}\sin^{2}\Delta_{21} \right)\\
		& - 4\left|U_{e3}\right|^{2}\left(1-\left|U_{e3}\right|^{2}\right)\sin^{2}\Delta_{3i} \\
		= & P_{ee}^{3\nu} - 4\left|U_{e0}\right|^{2}\left[ \left|U_{e1}\right|^{2}\sin^{2}\Delta_{01} + \left|U_{e2}\right|^{2}\left(\sin^{2}\Delta_{02}-\sin^{2}\Delta_{21}\right)\right]
		\end{aligned}
		\end{equation}
		
		\item $\Delta m_{01}^{2} \ll 10^{-9} \ \rm{eV^{2}}$.
		In this case, we can approximatively consider that $\sin^{2}\Delta_{01} \approx 0$ and $\Delta_{02} \approx \Delta_{21}$, then Eq.~(\ref{eq.kam_03=0}) is simplified into
		\begin{equation}
		\begin{aligned}
		\label{eq.kam_02=0_l}
		P_{ee}  
		\approx & 1 - 4\left(\left|U_{e0}\right|^{2}\left|U_{e2}\right|^{2} + \left|U_{e1}\right|^{2}\left|U_{e2}\right|^{2} \right)\sin^{2}\Delta_{21} - 4\left|U_{e3}\right|^{2}\left(1-\left|U_{e3}\right|^{2}\right)\sin^{2}\Delta_{3i} \\
		=       & P_{ee}^{3\nu}\,.
		\end{aligned}
		\end{equation}
	\end{enumerate}

	If only $\theta_{02}$ is nonzero for the sterile mixing angles, we have $\left|U_{e0}\right|^{2}=s^{2}_{02}\left|U_{e2}^{3\nu}\right|^{2}=s^{2}_{02}s^{2}_{12}c^{2}_{13}$, $\left|U_{e1}\right|^{2}=\left|U_{e1}^{3\nu}\right|^{2}=c^{2}_{12}c^{2}_{13}$, $\left|U_{e2}\right|^{2}=c^{2}_{02}\left|U_{e2}^{3\nu}\right|^{2}=c^{2}_{02}s^{2}_{12}c^{2}_{13}$ and $\left|U_{e3}\right|^{2}=\left|U_{e3}^{3\nu}\right|^{2}=s^{2}_{13}$; see Eq.~(\ref{eq.U_02}) in Appendix.~\ref{sec:Um}. Then,
	
	\begin{enumerate}
		\item $\Delta m_{01}^{2} \gg \Delta m_{21}^{2}$.
		Since the phase $\Delta_{01}$, $\Delta_{02}$ of the sine in Eq.~(\ref{eq.kam_03=0}) is so large, the survival probability can be similarly given by
		\begin{equation}
		\label{eq.kam_01=0_h}
		\begin{aligned}
		P_{ee}  
		\approx & 1 - 2\left( \left|U_{e0}\right|^{2}\left|U_{e1}\right|^{2} + \left|U_{e0}\right|^{2}\left|U_{e2}\right|^{2}\right) - 4\left(\left|U_{e1}\right|^{2}\left|U_{e2}\right|^{2}\sin^{2}\Delta_{21} \right) \\
		& - 4 \left|U_{e3}\right|^{2}\left(1-\left|U_{e3}\right|^{2}\right)\sin^{2}\Delta_{3i} \\
		= & P_{ee}^{3\nu} - 2\left|U_{e0}\right|^{2}\left(\left|U_{e1}\right|^{2} + \left|U_{e2}\right|^{2}\right) + 4\left(\left|U_{e0}\right|^{2}\left|U_{e1}\right|^{2}\sin^{2}\Delta_{21} \right)
		\end{aligned}
		\end{equation}
		
		\item $\Delta m_{01}^{2} \sim \Delta m_{21}^{2}$.
		In this case, Eq.~(\ref{eq.kam_03=0}) is simplified into
		\begin{equation}
		\label{eq.kam_01=0_m}
		\begin{aligned}
		P_{ee}  
		\approx & 1 - 4\left( \left|U_{e0}\right|^{2}\left|U_{e1}\right|^{2}\sin^{2}\Delta_{01} + \left|U_{e0}\right|^{2}\left|U_{e2}\right|^{2}\sin^{2}\Delta_{02} + \left|U_{e1}\right|^{2}\left|U_{e2}\right|^{2}\sin^{2}\Delta_{21} \right)\\
		& - 4 \left|U_{e3}\right|^{2}\left(1-\left|U_{e3}\right|^{2}\right)\sin^{2}\Delta_{3i} \\
		= & P_{ee}^{3\nu} - 4\left|U_{e0}\right|^{2}\left[ \left|U_{e2}\right|^{2}\sin^{2}\Delta_{02} + \left|U_{e1}\right|^{2}\left(\sin^{2}\Delta_{01}-\sin^{2}\Delta_{21}\right)\right]
		\end{aligned}
		\end{equation}
If $\Delta m_{01}^{2} \approx \Delta m_{21}^{2}$ is satisfied, one can approximatively consider that $\Delta_{02} \approx 0$ and $\Delta_{01} \approx \Delta_{21}$. As a result, we can get $P_{ee}\approx P_{ee}^{3\nu}$ from Eq.~(\ref{eq.kam_01=0_m}).
		
		\item $\Delta m_{01}^{2} \ll 10^{-9} \ \rm{eV^{2}}$.
		In this case, we make an approximation that $\sin^{2}\Delta_{01} \approx 0$ and $\Delta_{02} \approx \Delta_{21}$, then Eq.~(\ref{eq.kam_03=0}) becomes
		\begin{equation}
		\begin{aligned}
		\label{eq.kam_01=0_l}
		P_{ee}  
		\approx & 1 - 4\left(\left|U_{e0}\right|^{2}\left|U_{e2}\right|^{2} + \left|U_{e1}\right|^{2}\left|U_{e2}\right|^{2} \right)\sin^{2}\Delta_{21} - 4 \left|U_{e3}\right|^{2}\left(1-\left|U_{e3}\right|^{2}\right)\sin^{2}\Delta_{3i} \\
		\approx & P_{ee}^{3\nu} - 4\left|U_{e0}\right|^{2}\left(\left|U_{e2}\right|^{2} - \left|U_{e1}\right|^{2}\right)\sin^{2}\Delta_{21} \\
		\end{aligned}
		\end{equation}
	\end{enumerate}


\bibliographystyle{JHEP}
\bibliography{refs}
\end{document}